\documentclass[twocolumn,prl,aps,superscriptaddress,nofootinbib]{revtex4}

\usepackage{amsmath}
\usepackage{bm}
\usepackage{graphicx}
\usepackage{hyperref}
\usepackage{xspace}
\usepackage{amsfonts}

\newcommand{\mapPlanck}{\includegraphics[width=0.7cm,height=0.3cm,trim=-1cm 2cm 0
0]{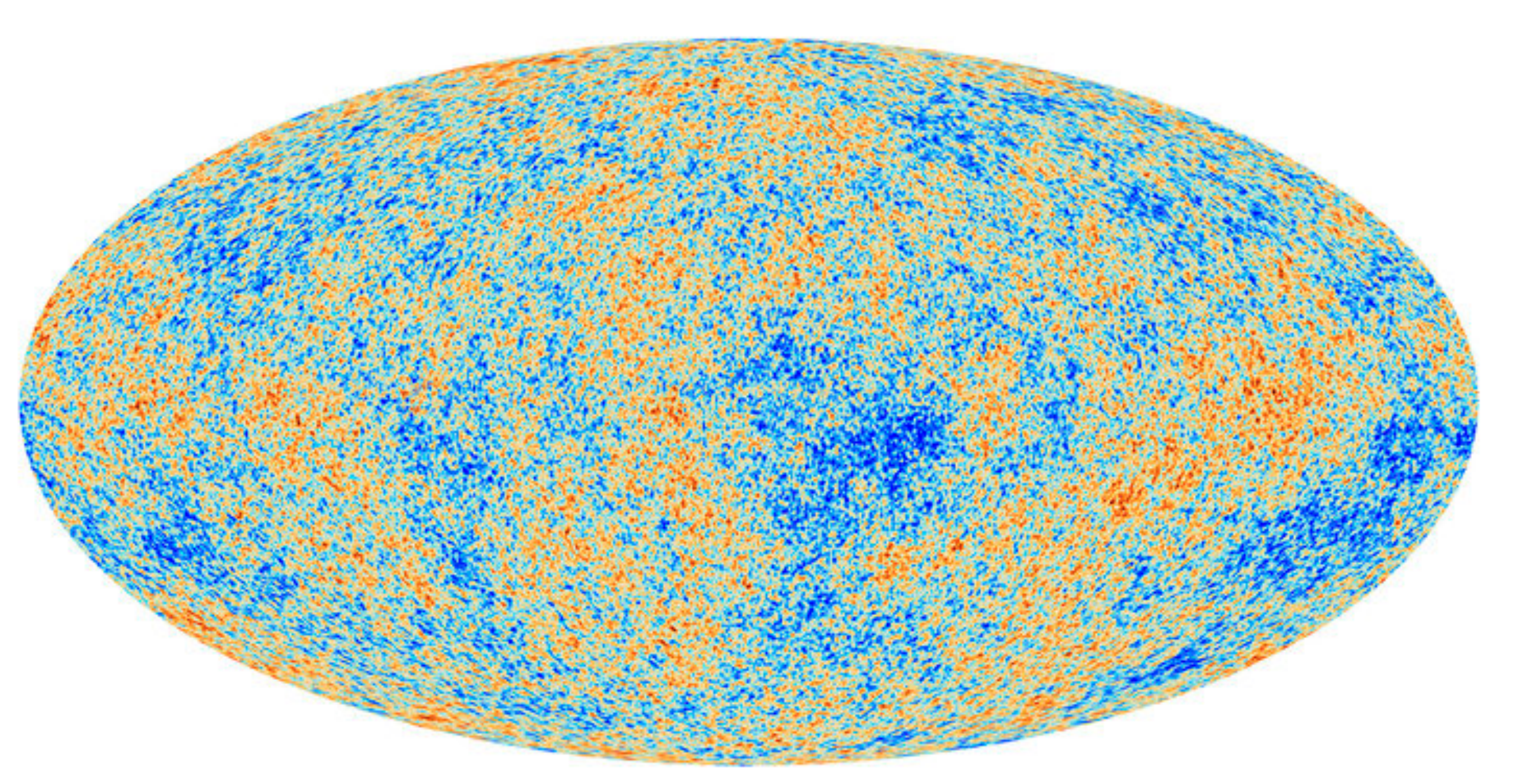}}
\newcommand{\ketmapPlanck}{\vert\mapPlanck\rangle}

\newcommand{\map}{\includegraphics[width=0.7cm,height=0.3cm,trim=-1cm 2cm 0
0]{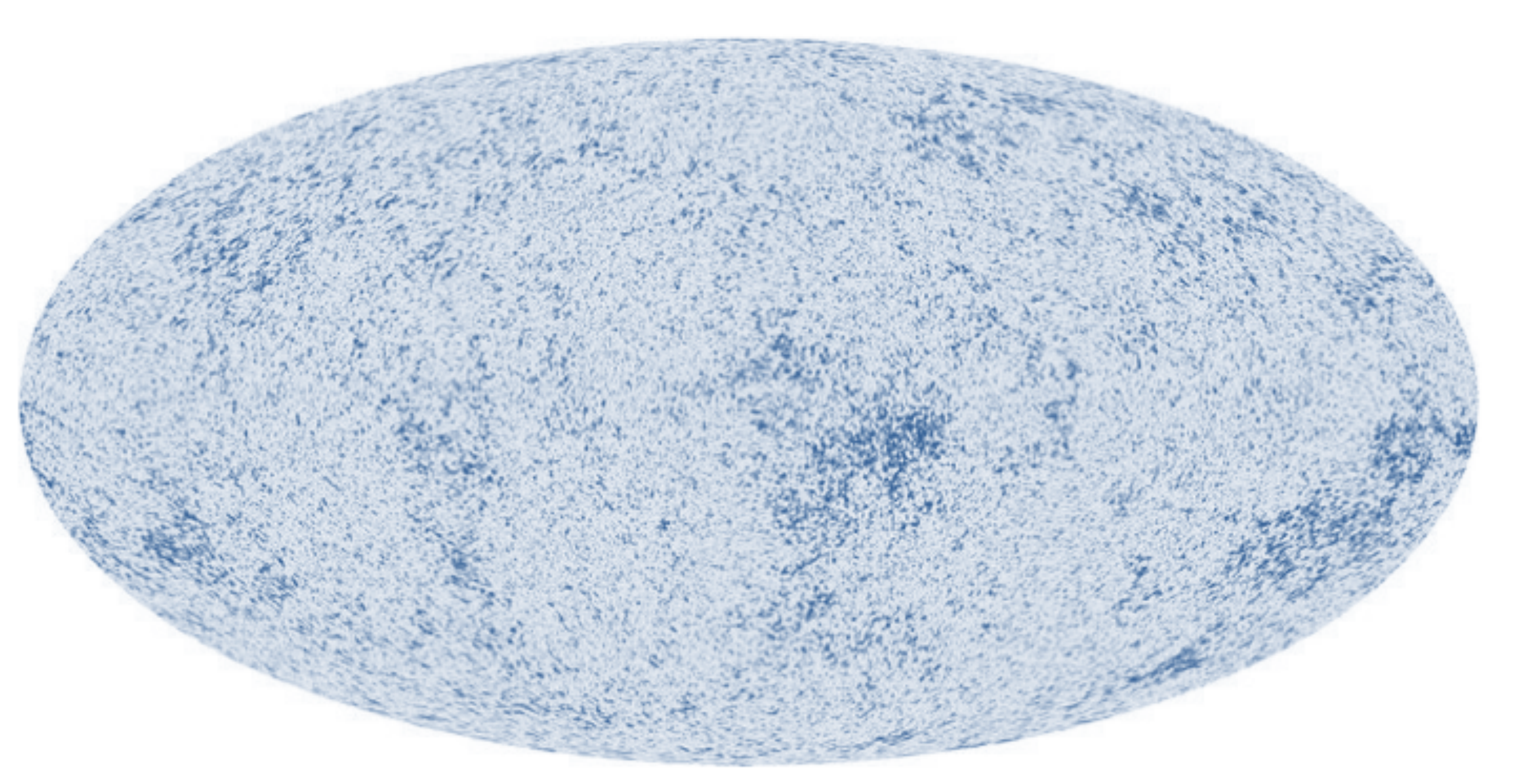}}
\newcommand{\ketmap}{\vert\map\rangle}

\makeatletter
\gdef\@fpheader{}
\g@addto@macro\bfseries{\boldmath}
\makeatother

\hypersetup{
    bookmarks=true,         
    unicode=false,          
    pdftoolbar=true,        
    pdfmenubar=true,        
    pdffitwindow=false,     
    pdfstartview={FitH},    
    pdfnewwindow=true,      
    colorlinks=true,        
    linkcolor=red,          
    citecolor=cyan,         
    filecolor=magenta,      
    urlcolor=blue,         
    linktocpage=true
}

\newcommand{\letterSec}[1]{\textit{#1}---}

\newcommand{\ie}{{i.e.~}}

\newcommand{\eg}{\textsl{e.g.~}}

\newcommand{\dd}{\mathrm{d}}
\newcommand{\ee}{e}

\newcommand{\sss}[1]{{\scriptscriptstyle{#1}}}

\newcommand{\uPl}{\mathrm{Pl}}

\newcommand{\umin}{\mathrm{min}}
\newcommand{\umax}{\mathrm{max}}
\newcommand{\uend}{\mathrm{end}}

\newcommand{\urad}{\mathrm{rad}}

\newcommand{\uc}{\mathrm{c}}

\newcommand{\uS}{\mathrm{S}}

\newcommand{\usssS}{\sss{\uS}}

\newcommand{\usssPl}{\sss{\uPl}}

\newcommand{\nS}{n_\usssS}

\newcommand{\setR}{\mathbb{R}}

\newcommand{\Rea}{\Re \mathrm{e}\,}
\newcommand{\Ima}{\Im \mathrm{m}\,}

\newcommand{\GeV}{\mathrm{GeV}}

\newcommand{\cs}{c_{_\mathrm{S}}}

\newcommand{\Mp}{M_\usssPl}

\newcommand{\beq}{\begin{equation}}
\newcommand{\eeq}{\end{equation}}
\newcommand{\bea}{\begin{eqnarray}}
\newcommand{\eea}{\end{eqnarray}}

\newlength{\wsingfig}
\newlength{\wdblefig}
\newlength{\wquadfig}
\newlength{\wtriplefig}
\newcommand{\Eq}[1]{Eq.~(\ref{#1})}
\newcommand{\Eqs}[1]{Eqs.~(\ref{#1})}
\newcommand{\Fig}[1]{Fig.~{\ref{#1}}}

\newcommand{\Refa}[1]{Ref.~{\cite{#1}}}

\begin{document}

\title{A cosmic shadow on CSL}

\author{J\'er\^ ome Martin} \email{jmartin@iap.fr}
\affiliation{Institut d'Astrophysique de Paris, UMR 7095-CNRS,
  Universit\'e Pierre et Marie Curie, 98bis boulevard Arago, 75014
  Paris, France}

\author{Vincent Vennin} \email{vincent.vennin@apc.in2p3.fr}
\affiliation{Laboratoire Astroparticule et Cosmologie, Universit\'e
  Denis Diderot Paris 7, 75013 Paris, France} \affiliation{Institut
  d'Astrophysique de Paris, UMR 7095-CNRS, Universit\'e Pierre et
  Marie Curie, 98bis boulevard Arago, 75014 Paris, France}

\date{\today}

\begin{abstract}
  The Continuous Spontaneous Localisation (CSL) model solves the
  measurement problem of standard quantum mechanics, by coupling the
  mass density of a quantum system to a white-noise field.  Since the
  mass density is not uniquely defined in general relativity, this
  model is ambiguous when applied to cosmology. We however show that
  most natural choices of the density contrast already make current
  measurements of the cosmic microwave background incompatible with
  other laboratory experiments.
\end{abstract}

\maketitle

%
%
Addressing the measurement (or macro-objectification) problem is a
central issue in quantum mechanics, and three classes of solutions
have been put forward~\cite{Bassi:2012bg}. One can either (1) leave
quantum theory unmodified and consider different interpretations (\eg
Copenhagen, many worlds, Qbism, {\it etc.}); (2) extend the
mathematical framework and introduce additional degrees of freedom
(\eg de Broglie-Bohm); or (3) consider that quantum theory is an
approximation of a more general framework and that, outside its domain
of validity, it differs from the standard formulation. Dynamical
collapse models~\cite{Ghirardi:1985mt, Diosi:1988uy, Ghirardi:1989cn,
  Bassi:2003gd, Bassi:2012bg} follow this last reasoning and introduce
a non-linear and stochastic modification to the Schr\"odinger
equation. Remarkably, the structure of this modification is
essentially unique. Through an embedded amplification mechanism, this
allows microscopic systems to be described by the standard rules of
quantum mechanics, while preventing macroscopic systems from being in
a superposition of macroscopically distinct configurations. It also
allows the Born rule to be derived rather than postulated~\cite{Bassi:2003gd}. Because
they lead to predictions that are different from that of conventional
quantum mechanics, dynamical collapse models are falsifiable contrary
to the other options mentioned before (except de Broglie-Bohm theory
in the out-of-equilibrium
regime~\cite{Valentini:1990zq,Valentini:1991fia}).

\begin{figure}[t]
\begin{center}
\includegraphics[width=0.49\textwidth]{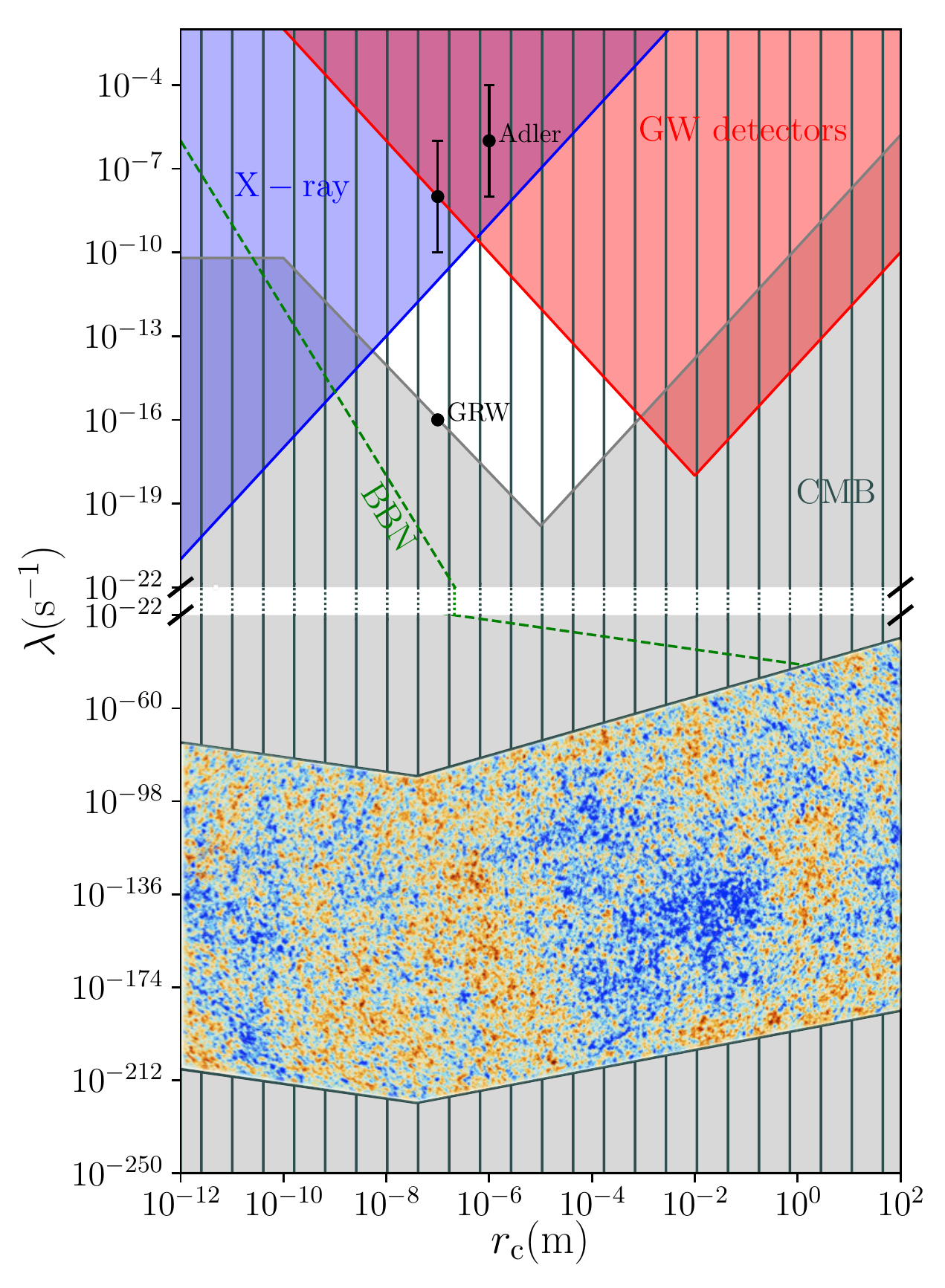}
\caption{Observational constraints on the two parameters $r_\uc$ and
  $\lambda$ of the CSL model. The white region is allowed by
  laboratory experiments while the unbarred region is allowed by CMB
  measurements (one uses $\Delta N=50$ for the pivot scale of the CMB,
  $H_{\mathrm{inf}}=10^{-5}\Mp$ and $\epsilon_1=0.005$). The two
  allowed regions are incompatible. The green dashed line stands for
  the upper bound on $\lambda$ if inflation proceeds at the Big-Bang
  Nucleosynthesis (BBN) scale.}
\label{fig:mapCSL}
\end{center}
\end{figure}

Different versions of dynamical collapse theories correspond to
different choices for the collapse operator (energy, momentum, spin,
position), the nature of the stochastic noise (white or non-white) and
whether dissipative effects are included or not. Only a collapse
operator related to position can ensure proper localisation in space,
and three iconic theories have been proposed: (1) the
Ghirardi-Rimini-Weber (GRW) model, which is historically the first one
but is not formulated in terms of a continuous stochastic differential
equation, (2) Quantum Mechanics with Universal Position Localisation
(QMUPL), where the collapse operator is position but where the
stochastic noise depends on time only, and (3) the Continuous
Spontaneous Localisation (CSL) model~\cite{Ghirardi:1989cn}, where the
stochastic noise depends on time and space and where the collapse
operator is the mass density. This version is the most refined of all
three, and features the modified Schr\"odinger equation
\begin{align}
\label{eq:cslphys}
 \kern-0.2em\dd \kern-0.2em\left\vert \Psi\right\rangle  \kern-0.2em & = 
 \kern-0.2em \biggl\lbrace \kern-0.2em -i \hat{H} \dd t + \kern-0.2em
\frac{\sqrt{\gamma}}{m_0} 
\displaystyle{\int}\kern-0.2em 
\dd \bm{x}_\mathrm{p}  \kern-0.1em\bigl[\hat{\rho}_{\mathrm{sm}} \kern-0.2em
\left(\bm{x}_\mathrm{p}\right)  \kern-0.2em
-  \kern-0.2em\left\langle \hat{\rho}_{\mathrm{sm}} \kern-0.2em
\left(\bm{x}_\mathrm{p}\right) 
\right\rangle \bigr]
\dd W_t \kern-0.2em\left(\bm{x}_\mathrm{p}\right)
\nonumber \\ &
-\frac{\gamma}{2m_0^2}\displaystyle{\int}\dd \bm{x}_\mathrm{p}
\left[\hat{\rho}_{\mathrm{sm}}
\left(\bm{x}_\mathrm{p}\right) 
- \left\langle \hat{\rho}_{\mathrm{sm}}
\kern-0.2em\left(\bm{x}_\mathrm{p}\right) 
\right\rangle \right]^2\dd t
\biggr\rbrace \left\vert \Psi\right\rangle ,
\end{align}
where $\hat{H}$ is the standard Hamiltonian of the system,
$\langle \hat{A} \rangle \equiv \langle \Psi \vert \hat{A} \vert
\Psi\rangle$,
$\gamma$ is the first free parameter of the theory, $m_0$ is a
reference mass (usually the mass of a nucleon),
$W_t(\bm{x}_\mathrm{p})$ is an ensemble of independent Wiener
processes (one for each point in space), and
$\hat{\rho}_{\mathrm{sm}}$ is the smeared mass density operator
\begin{align}
\label{eq:rho:cg}
\hat{\rho}_{\mathrm{sm}}\left(\bm{x}_\mathrm{p}\right) =  
\frac{1}{\left(2\pi\right)^{3/2} {r_\uc}^3} \int\dd\bm{y}_\mathrm{p} 
\, \hat{\rho}\left(\bm{x}_\mathrm{p}+\bm{y}_\mathrm{p}\right)
\ee^{-\frac{\left\vert \bm{y}_\mathrm{p}\right\vert^2}{2 r_\uc^2}},
\end{align}
where $ r_\uc$ is the second free parameter of the theory. The two
parameters $\gamma$ and $r_\uc$ have been constrained in various
laboratory experiments. The strongest bounds so far come from X-ray
spontaneous emission~\cite{2015arXiv150205961C}, force noise
measurements on ultracold cantilevers~\cite{2016PhRvL.116i0402V}, and
gravitational-wave interferometers~\cite{Carlesso:2016khv}. These
constraints leave the region of parameter space around
$r_\uc \sim 10^{-8}-10^{-4}\mathrm{m}$ and
$\lambda\sim 10^{-18}-10^{-10}\mathrm{s}^{-1}$ viable, where
$\lambda\equiv \gamma/(8\pi^{3/2}r_\uc^3)$, corresponding to the white
region in Fig.~\ref{fig:mapCSL}.

Dynamical collapse models can also be constrained in a cosmological
context~\cite{Perez:2005gh, Pearle:2007rw, Lochan:2012di,
  Martin:2012pea, Canate:2012ua, Piccirilli:2017mto, Leon:2017yna,
  Leon:2019jsl}. Indeed, the typical physical scales involved in
cosmology are many orders of magnitude different from those
encountered in the lab and this may lead to competitive constraints
(in the early universe, energy scales can be as high as $\sim
10^{15}\GeV$, corresponding to densities of $\sim 10^{80}\,
\mathrm{g}\times \mathrm{cm}^{-3}$). Moreover, one can argue that the
quantum measurement problem (as well as the quantum-to-classical
transition issue~\cite{PintoNeto:2011ui, Martin:2015qta,
  Martin:2017zxs, dePutter:2019xxv}) is even more acute in cosmology
than in the lab~\cite{Sudarsky:2009za}, due to the difficulties in
introducing an ``observer'' as in the standard Copenhagen
interpretation~\cite{1955mfqm.book.....V, Hartle:2019hae}.

Although the quantum state of cosmological perturbations, $\vert
\Psi_{2\, {\rm sq}}\rangle$, is a two-mode squeezed state that
features some classical properties~\cite{Polarski:1995jg,
  Albrecht:1992kf, Martin:2015qta}, it is not an eigenstate of the
Cosmic Microwave Background (CMB) temperature anisotropies, so how the
process
\begin{align}
\label{eq:collapse}
\vert \Psi_{2\, {\rm sq}}\rangle=
\sum_{ \map}c( \map)  \ketmap \rightarrow 
\ketmapPlanck_{\rm Planck}
\end{align}
occurred is unclear. This makes the early
universe a perfect arena to test CSL.
\begin{figure}[t]
\begin{center}
\includegraphics[width=0.45\textwidth]{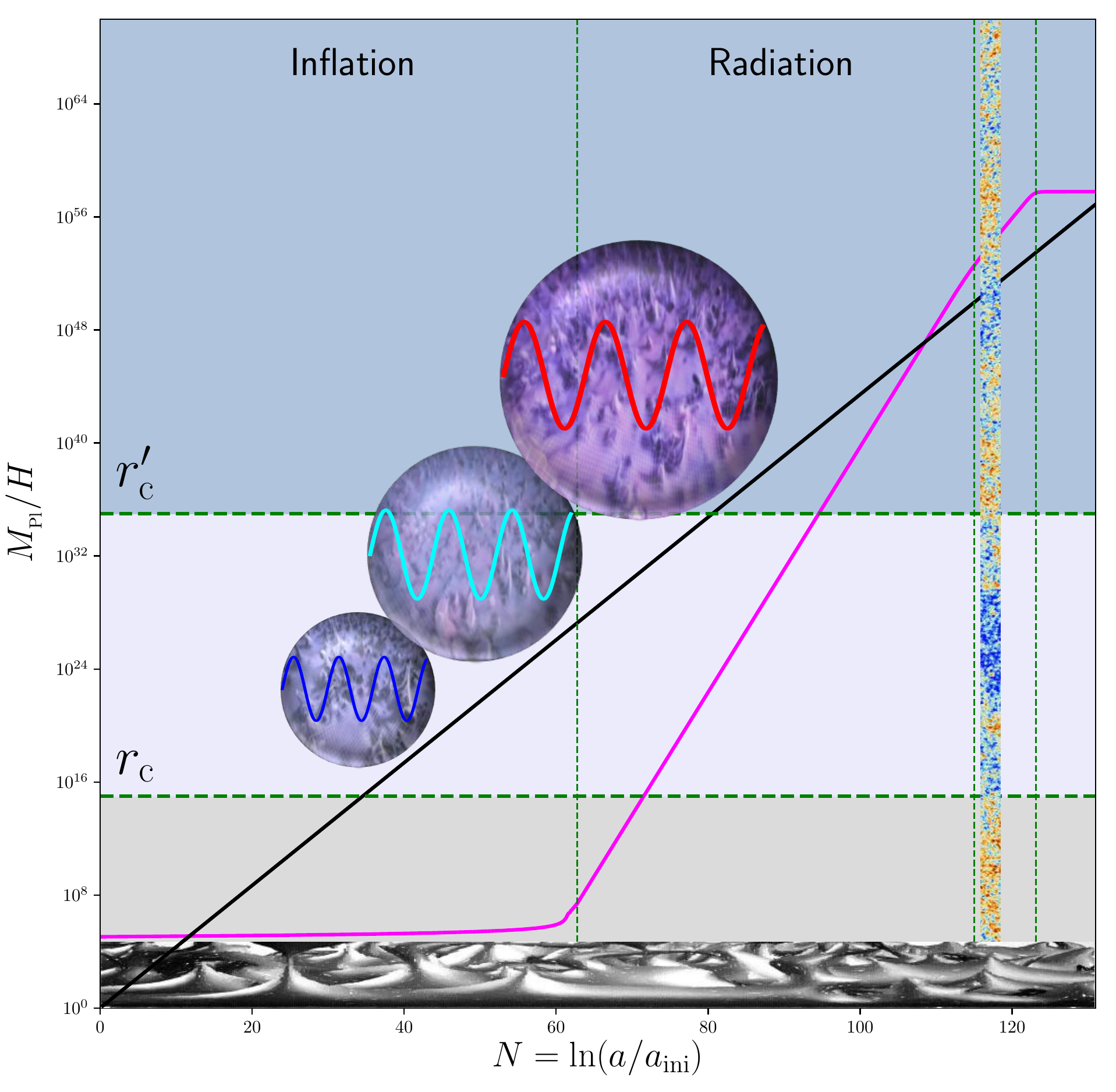}
\caption{Time evolution of the physical distances at play in the early
  universe. During inflation, the Hubble radius $H^{-1}$ (magenta
  line) is almost constant and, due to the expansion, the wavelength
  $\lambda_k$ of a Fourier mode (black line) for a given quantum field
  crosses out that scale, above which space-time curvature sources
  parametric amplification. In the subsequent Universe, $H^{-1}$
  increases faster than the scale factor $a$, hence $\lambda_k$
  crosses the Hubble radius back in. Depending on the value of
  $r_\uc$, $\lambda_k$ may cross out $r_\uc$ either during inflation
  ($r_\uc$) or during the radiation era ($r_\uc'$).}
\label{fig:CosmicSketch}
\end{center}
\end{figure}

The leading paradigm to describe this epoch is cosmic
inflation~\cite{Starobinsky:1980te,Guth:1980zm,Linde:1981mu,Albrecht:1982wi,Linde:1983gd},
which was introduced in order to solve the puzzles of the standard hot
big-bang phase. Inflation is believed to have been driven by a scalar
field $\phi$, named the ``inflaton'', the physical nature of which is
still unknown although detailed constraints on the shape of its
potential now
exist~\cite{Martin:2006rs,Lorenz:2007ze,Lorenz:2008je,Martin:2010hh,Martin:2010kz,Martin:2013tda,Martin:2013nzq,Martin:2014nya,Martin:2015dha}. Inflation
also provides a convincing mechanism for structure formation according
to which galaxies and CMB anisotropies are nothing but quantum vacuum
fluctuations amplified by gravitational instability and stretched to
astrophysical scales~\cite{Mukhanov:1981xt}. This mechanism fits very
well the high-accuracy astrophysical data now at our disposal, in
particular the CMB temperature and polarisation
anisotropies~\cite{Akrami:2018vks, Akrami:2018odb}.

The universe is well described by a flat, homogeneous and
isotropic metric of the Friedmann-Lema\^itre-Robertson-Walker (FLRW)
type, $\dd s^2=-\dd t^2+a^2(t)\delta_{ij}\dd x^i\dd x^j$, where $x^i$
is the comoving spatial coordinate, $t$ refers to cosmic time, and
$a(t)$ is the scale factor which depends on time only. During
inflation, the expansion is accelerated, $\ddot{a}>0$, and the Hubble
parameter $H=\dot{a}/a$ (where a dot denotes derivation with respect
to time) is almost constant, see \Fig{fig:CosmicSketch}. 

To describe the small quantum fluctuations living on top of this FLRW
background, the metric and inflaton fields are expanded according to
$g_{\mu \nu}=g_{\mu \nu}^{_{\rm FLRW}}(t)+\delta \hat{g}_{\mu
  \nu}(t,{\bm x})$
and $\phi=\phi^{_{\rm FLRW}}(t)+\delta \hat{\phi}(t,{\bm x})$ with
$\vert \delta g_{\mu \nu}/g_{\mu \nu}^{_{\rm FLRW}}\vert \ll 1$ and
$\vert \delta \phi/\phi^{_{\rm FLRW}}\vert \ll 1$. This gives rise to
two types of perturbations, scalars and tensors. Tensors correspond to
primordial gravitational waves and have not yet been detected, the
tensor-to-scalar ratio $r$ being
$r\lesssim 0.064$~\cite{Akrami:2018odb}. Then, scalar perturbations
can be described with a single gauge-invariant degree of freedom, the
so-called curvature perturbation
$\hat{\zeta}(t,{\bm x})$~\cite{Mukhanov:1981xt,Kodama:1985bj}, which
can be directly related to temperature anisotropies. Expanding the
action of the system (namely the Einstein-Hilbert action plus the
action of a scalar field) up to second order in the perturbations
leads to the Hamiltonian of the perturbations,
$\hat{H}=\int _{\setR^{3+}}\dd^3 {\bm k}\left[\hat{p}_{\bm
    k}^2+\omega^2(k,\eta)\hat{v}_{\bm k}^2\right]$,
where $\hat{v}_{\bm k} \equiv z \hat{\zeta}_{\bm k}$ is the
Mukhanov-Sasaki variable. One has introduced
$z\equiv a\sqrt{2\epsilon_1}\Mp/\cs$ where $\cs$ is the speed of sound
($\cs=1$ for a scalar field) and $\epsilon_1\equiv - \dot{H}/H^2$ is
the first Hubble-flow parameter~\cite{Schwarz:2001vv,Leach:2002ar}. In
the above expressions, the curvature perturbation has been Fourier
transformed,
$\hat{\zeta}(\eta,{\bm x})=(2\pi)^{-3/2} \int \dd ^3{\bm k}\,
\hat{\zeta}_{\bm k}(\eta)e^{i{\bm k}\cdot {\bm x}}$,
as appropriate for a linear theory where the modes evolve
independently. The conjugate momentum is
$\hat{p}_{\bm k}\equiv \hat{v}_{\bm k}'$, where a prime denotes
derivation with respect to the conformal time $\eta$ defined via
$\dd t= a \dd\eta$. Each mode behaves as a parametric oscillator,
$\hat{v}_{\bm k}''+\omega^2(k,\eta)\hat{v}_{\bm k}=0$, with a
time-dependent frequency $\omega^2(k,\eta)=\cs^2 k^2-z''/z$ that
involves the background dynamics. This phenomenon, described by the
interaction between a quantum field (here the cosmological
perturbations) and a time-dependent classical source (here the
background spacetime), leads to parametric amplification and can be
found in many other branches of Physics (\eg the Schwinger
effect~\cite{Schwinger:1951nm}, the dynamical Casimir
effect~\cite{Dodonov:2010zza}, Unruh~\cite{Unruh:1976db} and
Hawking~\cite{Hawking:1974sw} effects, {\it etc.}).

Quantisation of parametric oscillators yields squeezed states, which
are Gaussian states. Solving the Schr\"odinger equation with the above
Hamiltonian leads to
$\Psi[v]=\prod _{{\bm k},s}\Psi_{\bm k}^s(v_{\bm k}^s)$, where $s=$R,I
labels the real and imaginary parts of $v_{\bm k}$, with
$\Psi_{\bm k}^s(v_{\bm k}^s)=N_{\bm k}e^{-\Omega _{\bm k}(v_{\bm
    k}^s)^2}$,
$\vert N_{\bm{k}}\vert =\left(2\Rea \Omega_{\bm{k}}/\pi\right)^{1/4}$
and $\Omega_{\bm k}$ obeying the equation
$\Omega_{\bm k}'=-2i\Omega_{\bm k}^2+i\omega^2(k,\eta)/2$. In the
standard approach, $\langle \hat{v}_{\bm k}\rangle =0$ and one needs
to assume the existence of a specific process~(\ref{eq:collapse}) that
led to a particular realisation corresponding to our universe (this is
the macro-objectification problem mentioned above). The dispersion of
the different realisations is characterised by the two-point
correlation function
$\langle \zeta^2\rangle =\int {\cal P}_\zeta \dd \ln k $ where
${\cal P}_\zeta=k^3\vert \zeta_{\bm k}\vert^2/(2\pi^2)$ is the power
spectrum, which is predicted to be of the form $A_{_{\rm S}}k^{\nS-1}$
where $\nS$ should be close to one. The recent Planck data
(identifying spatial and ensemble averages) have confirmed this result
with $\ln \left(10^{10}A_{_{\rm S}}\right)=3.044\pm 0.014$ and
$n_{_{\rm S}}=0.9649\pm 0.0042$~\cite{Akrami:2018odb}.

If quantum theory is described by CSL rather than by the standard
framework, the behaviour of the cosmological perturbations is modified
according to \Eq{eq:cslphys}. In that case, the mass density is given
by $\rho=\bar{\rho}+\delta\rho$, where $\bar{\rho}$ is the homogeneous
component of the energy density satisfying the Friedmann equation
$\bar{\rho}=3\Mp^2H^2$, $\Mp$ is the reduced Planck mass, and $\delta\rho$ the
density fluctuation.

In General Relativity (GR) however, there is no unique definition of
the density contrast $\delta\rho/\bar{\rho}$. While all possible
choices coincide on sub-Hubble scales where observations are
performed, they can differ on super-Hubble scales. This introduces a
fundamental ambiguity when defining CSL in cosmology: each choice for
the density contrast leads to a different CSL theory.  In order to
illustrate how the calculation proceeds in details, we first consider
the physically well-motivated choice consisting in measuring the
energy density relative to the hypersurface which is as close as
possible to a ``Newtonian'' time slicing (denoted $\delta_\mathrm{g}$
in \Refa{Bardeen:1980kt}). This leads to $\delta\rho/\bar{\rho} =
-2\epsilon_1 \zeta +2\epsilon_1(1+3\epsilon_1a^2H^2\partial^{-2}) \zeta'
/(3aH)$ if the universe is dominated by a scalar field. Our aim is
certainly not to argue in favour of that specific choice, and at the
end of the paper we generalise our results to an arbitrary definition
of the density contrast.

From the previous considerations, \Eq{eq:cslphys} can be written in
Fourier space as a set of independent CSL equations for the real and
imaginary parts of each Fourier mode, in which the smeared mass
density operator reads
$\widehat{\delta\rho_{\mathrm{sm}}^{s}}\left(\bm{k}\right) =
\alpha_{\bm{k}}\hat{v}_{\bm{k}}^s+ \beta_{\bm{k}} \hat{p}_{\bm{k}}^s$
with
\begin{align}
\label{eq:alphainf}
\alpha_{\bm{k}} &\equiv \frac{ \Mp^2H^2 \epsilon_1}{z} \ee^{-\frac{k^2
    r_\uc^2}{2a^2}} \left[
-8 -\epsilon_2+
6\left(\frac{aH}{k}\right)^2\epsilon_1\left(1+\frac{\epsilon_2}{2}\right)\right] \\
\label{eq:betainf}
\beta_{\bm{k}} &\equiv  \frac{2\Mp^2H\epsilon_1}{az} \ee^{-\frac{k^2
    r_\uc^2}{2a^2}}\left[-3\epsilon_1\left(\frac{aH}{k}\right)^2+1\right],
\end{align}
where $\epsilon_2\equiv \dd\ln\epsilon_1/\dd\ln a$ denotes the second
Hubble-flow parameter. Because of the presence of the exponential term,
the effect of the CSL terms is triggered only once the mode $k$ under
consideration crosses out the scale $r_\uc$, \ie when its physical
wavelength is larger than $r_\uc$, $k/a< r_\uc^{-1}$. Depending on the
value of $r_\uc$, this can happen either during inflation or
subsequently, see \Fig{fig:CosmicSketch} (cases labeled $r_\uc$ and
$r_\uc'$, respectively). Physically, it is clear that the CSL terms
cannot ``localize'' a mode if its ``size'' (its wavelength) is smaller
than the localization scale $r_\uc$. This also means that, at early
time, when $k/a< r_\uc^{-1}$, the standard theory applies, which
implies that one of the great advantages of inflation, namely the
possibility to choose well-defined initial conditions in the Minkowski
limit (the so-called Bunch-Davies vacuum state~\cite{Bunch:1978yq}),
is preserved.

We are now in a position to solve \Eq{eq:cslphys}. The most general
stochastic Gaussian wavefunction can be written as
\begin{align}
\label{eq:stochawf}
\Psi_{\bm{k}}^s\left(v_{\bm{k}}^s\right)&= \vert
N_{\bm{k}}\left(\eta\right)\vert \exp\Bigl\lbrace -\Rea
\Omega_{\bm{k}}\left(\eta\right)
\left[v_{\bm{k}}^s-\bar{v}_{\bm{k}}^s\left(\eta\right)\right]^2
\nonumber \\ &
+i\sigma_{\bm k}^s(\eta)+i\chi_{\bm k}^s(\eta) v_{\bm{k}}^s -i\Ima
\Omega_{\bm k}(\eta ) \left(v_{\bm{k}}^s\right)^2\Bigr\rbrace,
\end{align}
where the free functions $\Omega_{\bm k}$, $\bar{v}_{\bm{k}}^s$,
$\sigma_{\bm k}^s$ and $\chi_{\bm k}^s$ are (a priori) stochastic
quantities. This wavepacket is centred around
$\left\langle \hat{v}_{\bm{k}}^s \right\rangle=\bar{v}_{\bm{k}}^s$
with a variance
$\left\langle \left(\hat{v}_{\bm{k}}^s-\bar{v}_{\bm{k}}^s\right)^2
\right\rangle = (4 \Rea \Omega_{\bm{k}} )^{-1}$.
The collapse of the
wavefunction happens if the width of $\Psi(v_{\bm k}^s)$ is much
smaller than the typical dispersion of its mean, \ie
\begin{align}
\label{eq:critere}
R\equiv \frac{\mathbb{E}\left[\left\langle 
\left(\hat{v}_{\bm k}^s-\bar{v}_{\bm k}^s\right)^2
\right \rangle\right]}{\mathbb{E}
\left(\bar{v}_{\bm k}^s{}^2\right)}\ll 1
\end{align}
where $\mathbb{E}$ denotes the stochastic average. In fact, if the
collapse occurs according to the Born rule, then
$\mathbb{E} \left(\bar{v}_{\bm k}^s{}^2\right)=\langle \hat{v}_{\bm
  k}^s{}^2\rangle_{\gamma=0}=(4 \Rea
\Omega_{\bm{k}}\vert_{\gamma=0})^{-1}$,
 and $R$ can also be defined as
$R=\mathbb{E}\left[\left\langle \left(\hat{v}_{\bm k}^s-\bar{v}_{\bm
        k}^s\right)^2 \right \rangle\right]/\langle \hat{v}_{\bm
  k}^s{}^2\rangle_{\gamma=0}$.

When the wavefunction has collapsed, its realisations are described by
$\bar{v}_{\bm{k}}^s$. The power spectrum of the Mukhanov-Sasaki
variable (or of curvature perturbation) is thus given by the
dispersion of that quantity,
\begin{align}
\label{eq:Pv:def}
{\cal P}_{v}\left(k\right) =\frac{k^3}{2\pi^2}
\left\{\mathbb{E} \left({{\bar{v}_{\bm{k}}^s}}{}^2\right)
-\left[\mathbb{E}\left({{\bar{v}_{\bm{k}}^s}}\right)\right]^2\right\}.
\end{align}
The above quantity can also be rewritten as
${\cal P}_{v}(k)=k^3\{\mathbb{E}(\langle \hat{v}_{\bm
  k}^s{}^2\rangle)-\mathbb{E}[\langle (\hat{v}_{\bm k}^s-\bar{v}_{\bm
  k}^s)^2\rangle]\}/(2\pi^2)$.

In order to calculate the quantities~(\ref{eq:critere})
and~(\ref{eq:Pv:def}), one can insert the stochastic
wavefunction~(\ref{eq:stochawf}) into \Eq{eq:cslphys} and solve the
obtained stochastic differential equations. One obtains that
$\Omega _{\bm k}$ decouples from the other free functions and obeys
$\Omega_{\bm{k}}'=4i\gamma a^4 \alpha_{\bm k}\beta _{\bm
  k}\Omega_{\bm{k}}/m_0^2 -2\left(i+2\gamma a^4 \beta_{\bm
    k}^2/m_0^2\right)\Omega_{\bm{k}}^2 + \gamma a^4\alpha_{\bm
  k}^2/m_0^2 + i\omega^2(k,\eta)/2$.
This equation is non-stochastic, as in the standard case, but contains
new terms proportional to $\gamma$. Since it is non-stochastic,
$\mathbb{E}[\langle (\hat{v}_{\bm k}^s-\bar{v}_{\bm
  k}^s)^2\rangle]=(4\Rea \Omega_{\bm{k}})^{-1} $
and this implies that
$R=\Rea \Omega_{\bm{k}}\vert_{\gamma=0}/\Rea \Omega_{\bm{k}}$.

In order to obtain the spectrum~(\ref{eq:Pv:def}),
$\mathbb{E}(\langle \hat{v}_{\bm k}^s{}^2\rangle)$ remains to be
determined. This is done by noticing that \Eq{eq:cslphys} can be
cast into a Lindblad equation~\cite{Lindblad:1975ef} for the averaged
density matrix
$\hat{\rho} =
\mathbb{E}(\vert\Psi\rangle\langle\Psi\vert)$~\cite{Bassi:2003gd}.
From this Lindblad equation, one can derive a third-order differential
equation for
$\mathbb{E} \left(\left\langle \hat{v}_{\bm{k}}^s{}^2
  \right\rangle\right)$
that can be solved exactly~\cite{Martin:2018zbe}. Combining the above mentioned results, one
obtains
\begin{align}
\label{eq:powerSpectrum:infl}
{\cal P}_{v}(k)&\simeq \frac{k^3}{2\pi^2}\frac{1}{4 \Rea 
\Omega_{\bm{k}}\vert_{\gamma=0}} 
\biggl[1+6 \frac{\gamma}{m_0^2}\epsilon_1^3\bar{\rho}_\mathrm{inf}
\left(\frac{k}{aH}\right)^{-1}_\uend
\nonumber \\ &
-\frac{\Rea \Omega_{\bm k}\vert_{\gamma=0}}{\Rea \Omega_{\bm{k}}}\biggr],
\end{align}
where $\bar{\rho}_\mathrm{inf}=3H_\mathrm{inf}^2\Mp^2$ is the energy
density during inflation.
\begin{figure}[t]
\begin{center}
\includegraphics[width=0.49\textwidth]{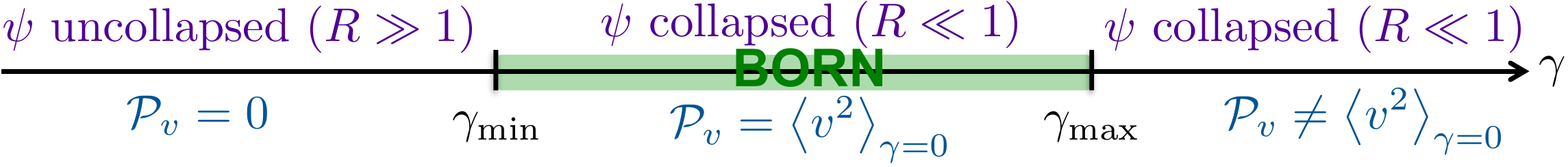}
\caption{Relevant values for $\gamma$. If $\gamma<\gamma_\umin$, the
  wavefunction does not collapse and the power spectrum vanishes. If
  $\gamma>\gamma_\umax$, the wavefunction collapses but the Born rule
  is violated and a non scale-invariant power spectrum is obtained,
  which is excluded by the CMB observations. The region where
  $\gamma_\umin<\gamma<\gamma_\umax$, unbarred in \Fig{fig:mapCSL}, is
  where the wavefunction collapses to a scale-invariant power
  spectrum.}
\label{fig:lambdaAxis}
\end{center}
\end{figure}
Depending on the value of $\gamma$, different results can be obtained,
that are sketched in \Fig{fig:lambdaAxis}. If $\gamma=0$, the state
remains homogeneous and isotropic, and the spectrum vanishes. Then,
when $\gamma$ increases above a certain threshold, collapse occurs
($R\ll 1$) so the third term in \Eq{eq:powerSpectrum:infl} can be
neglected. Provided the second term remains also negligible, the Born
rule is thus recovered, and a scale-invariant power spectrum is
obtained, in agreement with observations. Finally, when $\gamma$
continues to increase so as to make the second term large, the power
spectrum is no longer frozen on large scales and acquires a spectral
index $\nS=0$, which is excluded by CMB observations.

The amplitude of the correction to the power spectrum is proportional
to the energy density during inflation measured in units of the
reference mass, which is clearly huge and illustrates the potential of
cosmology to test the quantum theory, given that its characteristic
scales differ by orders of magnitude from those in the lab. The
correction is also slow-roll suppressed because of the relation
between $\delta\rho/\rho$ and $\zeta$ [since only the perturbations
are quantized, the classical part $\bar{\rho}$ cancels out in
\Eq{eq:cslphys}]. This suppression, however, is not sufficient to
compensate for the hugeness of $\bar{\rho}_{\rm inf}/m_0^2$.

In the standard situation, since the power spectrum of $\zeta$ is
frozen on large scales, its value at the end of inflation is what we
observe on the CMB last scattering surface and the calculation can be
stopped here. In the CSL theory however, this may no longer be true,
hence one needs to extend the present analysis to the radiation era
that follows inflation. During this epoch, the quantities
$\alpha_{\bm k}$ and $\beta_{\bm k}$ read
\begin{align}
\label{eq:alpharad}
\alpha_{\bm{k}} &\equiv \frac{24\Mp^2H^2}{z} \ee^{-\frac{k^2
    r_\uc^2}{2a^2}} \left[3 \left(\frac{aH}{k}\right)^2-1\right], \\
\label{eq:betarad}
\beta_{\bm{k}} &\equiv  \frac{12\Mp^2H}{az} \ee^{-\frac{k^2
    r_\uc^2}{2a^2}}\left[1-6\left(\frac{aH}{k}\right)^2\right].
\end{align}
The power spectrum of the Mukhanov-Sasaki variable can then be
determined using the same techniques as before, and one obtains
\begin{align}
{\cal P}_{v}(k)&=\frac{k^3}{2\pi^2}
\frac{1}{4\Rea \Omega_{\bm k}\vert_{\gamma=0}}
\biggl[1
\nonumber \\ &
+\frac{448}{3}\frac{\gamma}{m_0^2}
\bar{\rho}_\uend \epsilon_1 \left(\frac{k}{aH}\right)^{-1}_\uend
-\frac{\Rea \Omega_{\bm k}\vert_{\gamma=0}}{\Rea \Omega_{\bm k}}
\biggr],
\label{eq:powerSpectrum:inf}
\end{align}
where $\bar{\rho}_\uend$ is the energy density at the end of
inflation. Comparing with \Eq{eq:powerSpectrum:infl}, one can see that
the power spectrum indeed evolves during the transition between
inflation and the radiation era, but quickly settles to a constant
value, which is therefore the power spectrum probed by CMB
experiments. The CSL terms introduce a correction with a spectral
index $\nS=0$. One can also determine the collapse criterion
$R$, and one finds $1/R-1=1152\gamma\bar{\rho}_\uend (-k\eta_\uend)^{-7}/m_0^2$.

So far, we have assumed that the scale $r_\uc$ was crossed out during
inflation. Let us now examine the situation where $r_\uc$ is crossed
out during the radiation era. In that case, prior to crossing and in
particular during the entire inflationary phase, the standard results
remain valid. After crossing, the CSL terms become important and,
using again the same techniques, one obtains
\begin{align}
{\cal P}_{v}(k)&=\frac{k^3}{2\pi^2}
\frac{1}{4\Rea \Omega_{\bm k}\vert_{\gamma=0}}
\biggl[1
+\frac{35408}{429}\frac{\gamma}{m_0^2}
\bar{\rho}_\uend\epsilon_1
\nonumber \\ & \times
\left(\frac{r_\uc}{\ell_{_{\rm H}}}\right)_{\rm end}^{-9} 
\left(\frac{k}{aH}\right)^{-10}_\uend
-\frac{\Rea \Omega_{\bm k}\vert_{\gamma=0}}{\Rea \Omega_{\bm k}}
\biggr].
\label{eq:powerSpectrum:rad}
\end{align}
As before, the spectrum is frozen out on super-Hubble scales, but the
CSL correction now has spectral index $\nS=-9$. The collapse
criterion is given by
$1/R-1=7264\gamma/(11m_0^2)\bar{\rho}_\uend (k\eta_\uend)^{-14}(H_\uend
r_\uc)^{-7}$.

Since the CSL corrections are strongly scale dependent, they are ruled
out by CMB measurements. Therefore, using that
$k/(aH)\vert_{\uend}=e^{-\Delta N}$, where $\Delta N$ is the number of
e-folds spent by a mode between Hubble radius crossing during
inflation and the end of inflation (typically, for scales of
cosmological interest today, $\Delta N\sim 50$), one concludes that
$\gamma \ll m_0^2 (448\bar{\rho}_\uend \epsilon_1 /3)^{-1}e^{-\Delta N}$
if $H_\uend r_\uc<e^{\Delta N}$ and
$\gamma \ll m_0^2 (35408\bar{\rho}_\uend\epsilon_1/429)^{-1}(H_\uend
r_\uc)^9 e^{-10\Delta N}$
if $H_\uend r_\uc>e^{\Delta N}$. Moreover, the requirements that
collapse has occurred when the CMB is emitted, which is equivalent to
$R<1$, leads to
$\gamma>m_0^2(1152\bar{\rho}_\uend)^{-1}(-k\eta_\uend)^7$ if
$H_\uend r_\uc<e^{\Delta N}$ and
$\gamma>m_0^2(7264\bar{\rho}_\uend/11)^{-1}(-k\eta_\uend)^{14}(H_\uend
r_\uc)^7$
if $H_\uend r_\uc>e^{\Delta N}$. These constraints are represented in
\Fig{fig:mapCSL}.

These results allow us to conclude that if the CSL theory is embedded
in GR with the ``Newtonian'' density contrast, then the parameter
values that remain allowed by current laboratory experiments are
excluded by CMB measurements. Therefore, that version of CSL is now
ruled out. As stressed above, other choices for the density contrast
could be made. On large scales, they can be generically related to the
Newtonian density contrast $\delta_\mathrm{g}$ by $\delta_p \propto
\delta_\mathrm{g} [k/(aH)]^p$, where $p$ is a free index. Then, the
term $\propto k^{-1}$ in \Eq{eq:powerSpectrum:inf} becomes $\propto
k^{2p-1}$, while in \Eq{eq:powerSpectrum:rad}, the term $\propto
k^{-10}$ becomes $\propto k^{4p-10}$ and the term $\propto
(r_\uc/\ell_H)^{-9}$ becomes $\propto (r_\uc/\ell_H)^{2p-9}$. This
implies that any choice corresponding to $p<2$ is ruled out. When
derived from a more fundamental theory, the CSL model should thus come
with a prescription for the density contrast, that crucially
conditions the cosmological constraints. However, as explained in the
supplementary material, any ``natural'' choice for the density
contrast leads to $p=0$, with the one exception of the density
contrast denoted $\delta_\mathrm{m}$ in \Refa{Bardeen:1980kt}, which
corresponds to $p=2$. Our result therefore demonstrates that
astrophysical data are already accurate enough to rule out CSL
theories, except for a small subset of choices for the density
contrast.

Further subtleties could also arise if the CSL model was formulated in
a field-theoretic
manner~\cite{Ghirardi:1989cn,Tumulka:2005ki,Bedingham:2010hz,2014JSP...154..623B}
(which is in principle required in the present context -- although at
linear order all Fourier modes decouple and can be treated
quantum-mechanically), where parameter values may \eg run with the
energy scale at which the experiment is performed. Other approaches,
\eg Di\'osi-Penrose model~\cite{Diosi:1988uy,Penrose:1996cv} where
gravity is responsible for the collapse or scenarios where dissipative
effects are taken into account~\cite{2015NatSR...512518S}, could also
lead to different results. Other scenarios for forming cosmological
structures in the early universe, such as bouncing cosmologies, could
also be investigated.

Despite these uncertainties, the fact that
astrophysical data can constrain CSL highlights the
usefulness of early universe observations to discuss foundational
issues in quantum mechanics.

\letterSec{Acknowledgments} V.V. acknowledges funding from the
European Union's Horizon 2020 research and innovation programme under
the Marie Sk\l odowska-Curie grant agreement N${}^0$ 750491. It is a
pleasure to thank Angelo Bassi for
interesting comments and discussions.

\begin{widetext}

\newpage

\section{\Large {{Supplementary Material}}}

\section{The CSL Master Equations}

The CSL equation is given by (see, for instance, Eq.~(4) of
\Refa{2015NatSR...512518S})
\begin{align}
\label{eq:cslphys:app}
\dd \left\vert \Psi\left[ v\right]\right\rangle &= 
\biggl\lbrace  - i \hat{H} \dd t + \frac{\sqrt{\gamma}}{m_0} 
\int\dd \bm{x}_\mathrm{p} \left[\hat{C}\left(\bm{x}_\mathrm{p}\right) 
- \left\langle \hat{C}\left(\bm{x}_\mathrm{p}\right) \right\rangle \right] 
\dd W_t\left(\bm{x}_\mathrm{p}\right)
\nonumber \\ &
-\frac{\gamma}{2m_0^2}\int\dd \bm{x}_\mathrm{p}\left[\hat{C}
\left(\bm{x}_\mathrm{p}\right) 
- \left\langle \hat{C}\left(\bm{x}_\mathrm{p}\right) 
\right\rangle \right]^2\dd t
\biggr\rbrace \left\vert \Psi\left[ v \right]\right\rangle\, ,
\end{align}
where $\gamma$ is a free parameter, $m_0$ a reference mass (usually
the mass of a nucleon), $\hat{H}$ the Hamiltonian of the system,
$\hat{C}$ the collapse operator and $W_t(\bm{x}_\mathrm{p})$ is an ensemble of
independent Wiener processes satisfying
$\mathbb{E} \left[\dd W_t({\bm x}_\mathrm{p}) \dd W_{t'}({\bm
    x}'_\mathrm{p})\right] =\delta ({\bm x}_{\rm p}-{\bm x}'_{\rm
  p})\delta (t-t')\dd t^2$.
This equation is written in physical coordinates ${\bm x}_\mathrm{p}$.
However, in cosmology, it is more convenient to work in terms of
comoving coordinates defined by ${\bm x}_\mathrm{p}=a{\bm x}$, where
$a$ is the time-dependent scale factor and describes how the size of the universe
evolves with time. Comoving coordinates are coordinates for which
the motion related to the expansion of the universe is subtracted
out. In terms of these coordinates, the CSL equation reads
\begin{align}
\label{eq:csl}
\dd \left\vert \Psi\left[v\right]\right\rangle &= 
\biggl\lbrace  - i \hat{H} \dd t + \frac{1}{m_0}\sqrt{\frac{\gamma}{a^3}} 
\int\dd \bm{x} \, a^3\left[\hat{C}\left(\bm{x}\right) 
- \left\langle \hat{C}\left(\bm{x}\right) \right\rangle \right] 
\dd W_t\left(\bm{x}\right)
\nonumber \\ &
-\frac{\gamma}{2m_0^2}\int\dd \bm{x} \, a^3\left[\hat{C}
\left(\bm{x}\right) 
- \left\langle \hat{C}\left(\bm{x}\right) 
\right\rangle \right]^2\dd t
\biggr\rbrace \left\vert \Psi\left[v\right]\right\rangle\, ,
\end{align}
with $\dd W_t({\bm x}_\mathrm{p})=a^{-3/2} \dd W_t({\bm x})$ and
$\mathbb{E}\left[\dd W_t({\bm x}) \dd W_{t'}({\bm x}')\right] =\delta
({\bm x}-{\bm x}')\delta (t-t')\dd t^2$, this last result coming 
from the fact that $\mathbb{E}\left[\dd W_t({\bm x}_\mathrm{p})
\dd W_{t'}({\bm x}'_\mathrm{p})\right]
=\delta (a{\bm x}-a{\bm x}')\delta (t-t') \dd t^2\nonumber 
=a^{-3}\delta ({\bm x}-{\bm x}')\delta (t-t')\dd t^2$.

Notice that other implementations of the spontaneous localization model have been considered~\cite{Perez:2005gh, Piccirilli:2017mto, Leon:2017yna, Leon:2019jsl}, where the collapse is phenomenologically described. In this framework, collapse instantaneously occurs on space-like hypersurfaces, when the wavelength of a given mode crosses out a certain threshold. In our case, the dynamics of the collapse is fully resolved, but it is interesting to notice that these effective implementations already found modifications to the scalar and tensor power spectra.

In the CSL theory, the collapse operator is taken to be the energy
density. Moreover, in cosmological perturbations theory, one writes
$\hat{\rho} = \bar{\rho}+\widehat{\delta\rho}$, where $\bar{\rho}$ is
the background energy density, and only the fluctuating part is
quantised. As a consequence, the classical background part does not
contribute to the CSL equation since
$\hat{C}({\bm x})-\langle
  \hat{C}\left(\bm{x}\right)\rangle=
\bar{\rho}+\widehat{\delta\rho}-\langle
  \bar{\rho}+\widehat{\delta\rho}\rangle=\widehat{\delta\rho}-\langle
\widehat{\delta\rho}\rangle$.
The collapse operator also needs to be coarse-grained over the
distance $r_\uc$, where $r_\uc$ is the other free parameter in the
model. One therefore introduces the Gaussian coarse-graining procedure
\begin{align}
f_{\mathrm{cg}}\left(\bm{x}\right) = \left(\frac{a}{r_\uc}\right)^3 
\frac{1}{\left(2\pi\right)^{3/2}} \int\dd\bm{y} 
f\left(\bm{x}+\bm{y}\right)
\ee^{-\frac{\left\vert \bm{y}\right\vert^2 a^2}{2 r_\uc^2}}\, .
\end{align}
This implies that the collapse operator used in the CSL equation reads
\begin{align}
\hat{C}\left(\bm{x}\right) = \bar{\rho} \left. 
\widehat{\frac{\delta\rho}{\bar{\rho}}}
\right\vert_{\mathrm{cg}}\left(\bm{x}\right)
=3 \Mp^2\frac{\mathcal{H}^2}{a^2} \left. 
\widehat{\frac{\delta\rho}{\bar \rho}}
\right\vert_{\mathrm{cg}}\left(\bm{x}\right),
\end{align}
where we have used the Friedmann equation relating $\mathcal{H}=a'/a$ to $\bar{\rho}$.

In cosmology, perturbation theory is usually formulated in Fourier
space. In the CSL context, this leads to one CSL equation for each
mode, namely~\cite{Martin:2019oqq}
\begin{align}
\dd \left\vert \Psi_{\bm{k}}^s\left(t\right)\right\rangle &= 
\biggl\lbrace  - i \hat{H}_{\bm{k}}^s \dd t + \frac{\sqrt{\gamma a^3}}{m_0} 
\left[\hat{C}^s\left(\bm{k}\right) - \left\langle \hat{C}^s
\left(\bm{k}\right) \right\rangle \right] \dd W_t^s({\bm k})
-\frac{\gamma  a^3}{2m_0^2}\left[\hat{C}^s\left(\bm{k}\right) 
- \left\langle \hat{C}^s\left(\bm{k}\right) 
\right\rangle \right]^2\dd t
\biggr\rbrace \left\vert \Psi_{\bm{k}}^s\left(t\right)\right\rangle\, ,
\end{align}
the index $s$ designating the real and imaginary parts,
$s=\mathrm{R},\mathrm{I}$. The correlation functions of the noise in 
Fourier space are given by
\begin{align}
\mathbb{E}\left[\dd W_t^{\mathrm{R}}({\bm k})\, 
\dd W_{t'}^{\mathrm{R}}({\bm k}')\right]
&=\mathbb{E}\left[\dd W_t^{\mathrm{I}}({\bm k})\, 
\dd W_{t'}^{\mathrm{I}}({\bm k}')\right]
=\delta({\bm k}-{\bm k}')\delta(t-t')\dd t^2,
\quad
\mathbb{E}\left[\dd W_t^{\mathrm{R}}({\bm k})\, 
\dd W_{t'}^{\mathrm{I}}({\bm k}')\right]
=0,
\end{align}
and the Fourier transform of the collapse operator reads
\begin{align}
\hat{C}\left(\bm{k}\right) & = 3 \Mp^2\frac{\mathcal{H}^2}{a^2}
\ee^{-\frac{k^2 r_\uc^2}{2a^2}} \widehat{\frac{\delta
    \rho}{\bar \rho}}\left(\bm{k}\right)\, .  
\end{align} 

The CSL equation can also be cast into a Lindblad equation, see for
instance Eq.~(21) of \Refa{2015NatSR...512518S}, which takes the
form
\begin{align}
\frac{\dd \hat{\rho}}{\dd t} = -i\left[\hat{{H}},\hat{\rho}\right]
-\frac{\gamma}{2m_0^2}\int\dd \bm{x} \, a^3\left[\hat{C}\left(\bm{x}\right),
\left[\hat{C}\left(\bm{x}\right),\hat{\rho}\right]\right]
\end{align}
for the mean density matrix
$\hat{\rho} = \mathbb{E}(\vert\Psi\rangle\langle\Psi\vert)$. In
Fourier space, this gives rise to one equation per Fourier mode, which
can be written as
\begin{align} 
\label{eq:lindbladfourier}
\frac{\dd
  \hat{\rho}_{\bm{k}}^s}{\dd t} =
-i\left[\hat{{H}}_{\bm{k}}^s,\hat{\rho}_{\bm{k}}^s\right]
-\frac{\gamma}{2m_0^2}a^3\left[\hat{C}^s\left(\bm{k}\right),
\left[\hat{C}^s\left(\bm{k}\right),
\hat{\rho}_{\bm{k}}^s\right]\right].
 \end{align} 
\section{Solving the Lindblad equation}
The stochastic mean of the quantum expectation value of some
observable $\hat{O}_{\bm{k}}^s$ is given by
$\mathbb{E}\left(\left\langle \hat{O}_{\bm{k}}^s\right\rangle\right) =
\mathrm{Tr}\left(\hat{\rho}_{\bm{k}}^s \hat{O}_{\bm{k}}^s\right)$,
where $\hat{\rho}_{\bm{k}}^s$ obeys \Eq{eq:lindbladfourier}. Differentiating this
expression with respect to conformal time (we
recall that conformal time $\eta$ is related to cosmic time $t$ by
$\dd t=a\dd \eta$) and making use of  \Eq{eq:lindbladfourier}, one
obtains
\begin{align}
\frac{\dd}{\dd \eta} \mathbb{E}\left(\left\langle \hat{O}_{\bm{k}}^s
\right\rangle\right) = \mathbb{E}\left(\left\langle 
\frac{\partial}{\partial \eta} \hat{O}_{\bm{k}}^s\right\rangle\right) 
- i \mathbb{E} \left(\left\langle \left[\hat{O}_{\bm{k}}^s,\hat{\mathcal{H}}_{\bm{k}}^s\right] 
\right\rangle\right)- \frac{\gamma a^4}{2m_0^2}\mathbb{E} \left(
\left[\left[\hat{O}_{\bm{k}}^s,\hat{C}_{\bm{k}}^s\right],
\hat{C}_{\bm{k}}^s\right]\right).
\end{align}
For one-point correlators, $\hat{O}_{\bm k}^s=v_{\bm k}^s$ and
$\hat{O}_{\bm k}^s=p_{\bm k}^s$, this gives rise to
\begin{align}
\label{eq:linearlinear}
\frac{\dd \mathbb{E} \left( \left\langle \hat{v}_{\bm k}^s\right\rangle\right)}{\dd  \eta } =
\mathbb{E} \left(\left\langle \hat{p}_{\bm k}^s \right\rangle \right)\, , \qquad 
\frac{\dd\mathbb{E} \left( \left\langle \hat{p}_{\bm k}^s \right\rangle\right)}{\dd  \eta } =
-\omega^2(k,\eta) \mathbb{E} \left(\left\langle \hat{v}_{\bm k}^s \right\rangle\right)\, ,
\end{align}
which is nothing but the Ehrenfest theorem. For two-point correlators, denoting $P_{vv}(k)=\mathbb{E} (\langle \hat{v}_{\bm k}^s{}^2 \rangle) $, $P_{pp}(k)=\mathbb{E} (\langle \hat{p}_{\bm k}^s{}^2 \rangle ) $, $P_{vp}(k)=\mathbb{E} (\langle \hat{v}_{\bm k}^s\hat{p}_{\bm k}^s \rangle) $ and $P_{pv}(k)=\mathbb{E} (\langle \hat{p}_{\bm k}^s\hat{v}_{\bm k}^s \rangle) $, one obtains
\begin{align}
\frac{\dd  P_{vv}(k)}{\dd  \eta}&=P_{vp}(k)+P_{pv}(k)
+\frac{\gamma}{m_0^2} a^4 \beta_{\bm k}^2, \\
\frac{\dd  \left[P_{vp}(k)+P_{pv}(k)\right]}{\dd  \eta} &
=2P_{pp}(k)-2w^2(k,\eta)P_{vv}(k)
- 2a^4\frac{\gamma}{m_0^2} \alpha_{\bm k} \beta_{\bm k}, \\
\frac{\dd  P_{pp}(k)}{\dd  \eta}&= -\omega^2(k,\eta)
\left[P_{pv}(k)+P_{vp}(k)\right]
+ a^4\frac{\gamma}{m_0^2} \alpha_{\bm k}^2\, ,
\end{align}
where the coefficients $\alpha_{\bm k}$ and $\beta_{\bm k}$ have been
defined in the main text, see \Eqs{eq:alphainf}-(\ref{eq:betainf}) and~(\ref{eq:alpharad})-(\ref{eq:betarad}). These
equations can be combined into a single third-order equation for
$P_{vv}$ only, which reads
\begin{align}
\label{eq:thirdvlinear}
\frac{\dd ^3P_{vv}}{\dd \eta^3}+4\omega^2(k,\eta)\frac{\dd P_{vv}}{\dd \eta}
+4 \omega \frac{\dd \omega}{\dd \eta}P_{vv}
=S\, ,
\end{align}
where $S$ is the source function given by 
\begin{align}
\label{eq:source:def}
S=\frac{\gamma}{m_0^2}\left[2 a^4 \left(\alpha_{\bm{k}}^2+\omega^2  \beta_{\bm{k}}^2\right)
-2\left(a^4 \alpha_{\bm{k}} \beta_{\bm{k}}\right)'
+\left(a^4\beta_{\bm{k}}^2\right)''\right] .
\end{align}
As we will show below, this source function encodes both the modifications to the power spectrum and the collapsing time. Let us note that it is invariant under phase-space canonical transforms, so the results derived hereafter would be the same if other canonical variables than $v_{\bm{k}}$ and $p_{\bm{k}}$ were used.

As shown in \Refa{Martin:2018zbe}, \Eq{eq:thirdvlinear} can be solved by introducing the Green function of the free theory,
\begin{align}
\label{eq:Green:def}
G(\eta,\bar{\eta})=\frac{1}{W}\left[g_{\bm k}^0{}^*(\bar{\eta})
g_{\bm k}^0(\eta)-g_{\bm k}^0(\bar{\eta})g_{\bm k}^0{}^*(\eta)\right]
\Theta\left(\eta-\bar{\eta}\right),
\end{align}
where $g_{\bm k}^0$ is a solution of the Mukhanov-Sasaki equation, $(g_{\bm k}^0)''+\omega^2(k,\eta)g_{\bm k}^0=0$, $W=g_{\bm k}^0{}'g_{\bm k}^0{}^*-g_{\bm k}^0g_{\bm k}^0{}^*{}'$ is its Wronskian, and where $\Theta(x)=1$ if $x\geq 0$ and $0$ otherwise is the Heaviside function. By construction, given the mode equation obeyed by $g_{\bm k}^0$, it is a constant. Then, the solution to \Eq{eq:thirdvlinear} reads
\begin{align}
\label{eq:exactsolquadratic:generic} 
P_{vv}(k) = g_{\bm k}^0\left(\eta\right)
g^0_{\bm k}{}^*\left(\eta\right) +\frac{1}{2}\int_{-\infty}^\eta S
\left(\bar{\eta}\right)
G^2(\eta,\bar{\eta})
\dd\bar{\eta}\, .
\end{align}
\subsection{Inflation}
During inflation $a\simeq -1/(H\eta)$, and at leading order in the Hubble-flow parameters, \Eq{eq:source:def} gives rise to
\begin{align}
\label{eq:sourceinf}
S_{\mathrm{inf}}
&\simeq \frac{4\gamma}{m_0^2} \epsilon_1 H^2\Mp^2 k^2
\ee^{-\left(r_\uc/\lambda\right)^2} 
\left(\frac{\ell_{_{\rm H}}}{\lambda}\right)^{-6}
\biggl[126\epsilon_1^2-75 \epsilon_1 
\left(\frac{\ell_{_{\rm H}}}{\lambda}\right)^{2}
+81\epsilon_1^2\left(\frac{r_\uc}{\lambda}\right)^2
+18\left(\frac{\ell_{_{\rm H}}}{\lambda}\right)^{4}
-48\epsilon_1\left(\frac{\ell_{_{\rm H}}}{\lambda}\right)^{2}
\left(\frac{r_\uc}{\lambda}\right)^2
\nonumber \\ &
+18\epsilon_1^2\left(\frac{r_\uc}{\lambda}\right)^4
+\left(\frac{\ell_{_{\rm H}}}{\lambda}\right)^{6}
+7\left(\frac{\ell_{_{\rm H}}}{\lambda}\right)^{4}
\left(\frac{r_\uc}{\lambda}\right)^2
-12\left(\frac{\ell_{_{\rm H}}}{\lambda}\right)^{2}
\left(\frac{r_\uc}{\lambda}\right)^4
+2\left(\frac{\ell_{_{\rm H}}}{\lambda}\right)^{4}
\left(\frac{r_\uc}{\lambda}\right)^4\biggr],
\end{align}
where $\ell_{_{\rm H}}=H^{-1}$ is the Hubble radius and
$\lambda =a(\eta)/k$ the wavelength of the Fourier mode with comoving
wavenumber $k$. The quantity $\ell_{_{\rm H}}/\lambda$ can also be
written as $\ell_{_{\rm H}}/\lambda=k/(aH)=-k\eta$. We see that the
amplitude of the source is controlled by the energy density during
inflation, $\bar{\rho}_{\rm inf}=3H^2\Mp^2$, and by the first Hubble-flow
parameter $\epsilon_1$ (at next-to-leading order in slow roll, higher-order Hubble flow parameters would appear). The limits we are
interested in are $\ell_{_{\rm H}}/\lambda\ll 1$ (super Hubble limit)
and $r_\uc/\lambda \ll 1$ (otherwise the exponential term turns the
source off, see the discussion in the main text). In this regime, the dominant term is the first one, proportional to $126\epsilon_1^2$ (although it is slow-roll suppressed). 

Normalising the mode function in the Bunch-Davies vacuum, at leading order in slow roll one has 
\bea
g_{\bm k}^0(\eta)=\frac{e^{ik\eta}}{\sqrt{2k}}\left(1+\frac{i}{k\eta}\right),
\eea
from which \Eq{eq:Green:def} gives
\begin{align}
\label{eq:Green:inf}
G_{\mathrm{inf}}(\eta,\bar{\eta}) = 
\frac{\left(1+k^2\eta \bar{\eta}\right)
\sin\left[k\left(\eta- \bar{\eta}\right)\right]-k\left(\eta- \bar{\eta}\right)\cos\left[k\left(\eta- \bar{\eta}\right)\right]}{k^3\eta \bar{\eta}}\Theta\left(\eta-\bar{\eta}\right)
\simeq
\frac{\eta^3-\bar{\eta}^3}{3\eta\bar{\eta}}
\Theta\left(\eta-\bar{\eta}\right)
\, .
\end{align}
The second expression is valid in the super-Hubble limits $-k\eta \rightarrow 0$ (since the power spectrum is computed on super-Hubble scales)
and $-k \bar{\eta} \rightarrow 0$ (since we assume $Hr_\uc\gg 1$, so any mode is super Hubble when it crosses out $r_\uc$). Plugging \Eqs{eq:sourceinf} and~(\ref{eq:Green:inf}) into \Eq{eq:exactsolquadratic:generic}, one obtains at leading order
\begin{align}
P_{vv}(k) &\simeq  
\vert v_{\bm k}\vert^2_{\rm standard}+\frac{18\gamma}{m_0^2k}H^2\Mp^2\epsilon_1^3
\left(\frac{k}{aH}\right)^{-3}=\vert v_{\bm k}\vert^2_{\rm standard}
\left[1+36\frac{\gamma}{m_0^2} H^2\Mp^2\epsilon_1^3
\left(\frac{k}{aH}\right)^{-1}\right],
\end{align}
where $\vert v_{\bm k}\vert^2_{\rm standard}=\vert g_{\bm k}^0\vert^2$,
which is the result used in the main text.
\subsection{Radiation-dominated epoch}
Let us now study what happens during the radiation dominated era. In
that case the scale factor is given by
$a(\eta)=a_{\mathrm{r}}\left(\eta-\eta_r\right)$ and, as a
consequence,
$\mathcal{H}(\eta)=a'/a=(\eta-\eta_{\mathrm{r}})^{-1}$. Requiring the
scale factor and its derivative (or, equivalently, the Hubble
parameter) to be continuous, which is equivalent to the continuity of
the first and second fundamental forms, gives
$\eta_{\mathrm{r}} = 2 \eta_\uend$ and
$a_{\mathrm{r}} = 1/(H_\uend \eta_\uend^2)$. 

Using the coefficients $\alpha_{\bm k}$ and
$\beta _{\bm k}$ given in \Eqs{eq:alpharad}
and~(\ref{eq:betarad}), the source function~(\ref{eq:source:def}) reads
\begin{align}
\label{eq:sourcerad}
 \nonumber
S_{\rm rad} 
&=8\frac{\gamma}{m_0^2} H_\uend^2\Mp^2 k^2\ee^{-\left(r_\uc/\lambda\right)^2} 
\left(\frac{a_\uend}{a}\right)^4
\left(\frac{\ell_{_{\rm H}}}{\lambda}\right)^{-6}
\biggl[3024-414\left(\frac{\ell_{_{\rm H}}}{\lambda}\right)^2
+\left(\frac{\ell_{_{\rm H}}}{\lambda}\right)^6
-1836\left(\frac{a_\uend}{a}\right)^2
\left(\frac{r_\uc}{\lambda}\right)^2_\uend
\\ \nonumber &
+216\left(\frac{a_\uend}{a}\right)^4
\left(\frac{r_\uc}{\lambda}\right)^4_\uend
-72\left(\frac{a_\uend}{a}\right)^4
\left(\frac{\ell_{_{\rm H}}}{\lambda}\right)^2
\left(\frac{r_\uc}{\lambda}\right)^4_\uend
+432\left(\frac{a_\uend}{a}\right)^2
\left(\frac{\ell_{_{\rm H}}}{\lambda}\right)^2
\left(\frac{r_\uc}{\lambda}\right)^2_\uend
\\ &
+6\left(\frac{a_\uend}{a}\right)^2
\left(\frac{\ell_{_{\rm H}}}{\lambda}\right)^2
\left(\frac{r_\uc}{\lambda}\right)^4_\uend
-21\left(\frac{a_\uend}{a}\right)^2
\left(\frac{\ell_{_{\rm H}}}{\lambda}\right)^4
\left(\frac{r_\uc}{\lambda}\right)^2_\uend
\biggr].
\end{align}
Its form is similar to that of the source during inflation, see \Eq{eq:sourceinf}, although the
amplitude is now proportional to the energy density at the end of
inflation, $\bar{\rho}_\uend=3H_\uend^2\Mp^2$, and is no longer slow-roll
suppressed as is expected in the radiation-dominated era. The
coefficients of the expansion depend on
$\left(r_\uc/\lambda\right)_\uend$, the ratio between the CSL scale and the mode wavelength
evaluated at the end of inflation. This dependence on quantities
evaluated at the end of inflation comes from the matching procedure.

At the perturbative
level, the Mukhanov-Sasaki variable now obeys
${g_{\bm k}^0}''+(c_{_{\rm S}}^2k^2-z''/z)g_{\bm k}^0=0$ with
$c_{_{\rm S}}^2=1/3$ and
$z= a\Mp \sqrt{2 \epsilon_1}/\cs = 2\sqrt{3}a\Mp$. The solution reads
\begin{align}
g_{\bm k}^0(\eta)= A_{\bm k} \ee^{-ik\frac{\eta-\eta_{\mathrm{r}}}{\sqrt{3}}} + B_{\bm k}  \ee^{ik\frac{\eta-\eta_{\mathrm{r}}}{\sqrt{3}}}\, ,
\end{align}
On super-Hubble scales, continuity of the first and second fundamental forms is
equivalent to the continuity of $\zeta$ and the Bardeen potential $\Phi$. At leading order in $k \eta_\uend$, this leads to
\begin{align} 
\label{eq:free:mode:rad}
g_{\bm k}^0(\eta) =
-\frac{3i}{\sqrt{k \epsilon_1}(k\eta_\uend)^2}
\sin\left[\frac{k}{\sqrt{3}}\left(\eta-
    \eta_{\mathrm{r}}\right)\right] .
\end{align}
Plugging this expression into \Eq{eq:Green:def}, one obtains 
\begin{align}
\label{eq:Green:rad}
G_\urad(\eta,\bar{\eta})=\frac{\sqrt{3}}{k}\sin\left[\frac{k}{\sqrt{3}}\left(\eta-\bar{\eta}\right)\right]\Theta\left(\eta-\bar{\eta}\right)\simeq \left(\eta-\bar{\eta}\right)\Theta\left(\eta-\bar{\eta}\right). 
\end{align}
At this stage, one must distinguish between two situations: either the
Fourier mode under consideration crosses out the scale $r_\uc$ during
inflation or during the radiation-dominated era. 
\subsubsection{Case where the mode crosses out $r_\uc$ during inflation}
In the standard situation, the power spectrum of $\zeta$ computed at the
end of inflation is frozen on super Hubble scales and can be directly
propagated to the last scattering surface. Here, however, a priori,
the power spectrum continues to evolve during the radiation-dominated
era even on large scales. 

The integral appearing in \Eq{eq:exactsolquadratic:generic} can be split in two parts: one for which $-\infty<\bar{\eta}<\eta_\uend$, which was already calculated above during inflation, and one for which $\eta_\uend<\bar{\eta}<\eta$ that we now calculate. If the
scale $r_\uc$ is crossed out during inflation, then
$(r_\uc/\lambda)_\uend\ll 1$ and all the terms in the source but the
one proportional to $3024$ can be ignored. At leading order in $\ell_{\mathrm{H}}/\lambda_\uend$ and $r_\uc/\lambda_\uend$, one obtains that, after a few $e$-folds, the power spectrum freezes to 
\begin{align}
P_{vv}(k)=\left \vert v_{\bm k}^s
\right \vert^2_{\rm standard}\left[1+448\frac{\gamma}{m_0^2}
H_\uend^2\Mp^2 \epsilon_1 \left(\frac{k}{aH}\right)^{-1}_\uend\right].
\end{align}
\subsubsection{Case where the mode crosses out $r_\uc$ during the radiation-dominated era}
The mode crosses out $r_\uc$ when $a_{\mathrm{cross}}=kr_\uc$, \ie at
$\eta_{\mathrm{cross}}=\eta_\mathrm{r}+k\eta_\uend^2H_\uend r_\uc$, which implies that
$(a_\uend/a_{\mathrm{cross}})(r_\uc/\lambda)_\uend=1$. As a
consequence, in the source~(\ref{eq:sourcerad}), the terms proportional to $3024$, $-1836$
and $216$ are of the same order of magnitude initially, while the
others are negligible since suppressed by powers of
$\ell_{_{\rm H}}/\lambda$ and can be safely neglected. This gives rise to
\begin{align}
P_{vv}(k)=\left \vert v_{\bm k}^s
\right \vert^2_{\rm standard}\left[1+\frac{35408}{143}
\frac{\gamma}{m_0^2}
H_\uend^2\Mp^2 \epsilon_1 \left(\frac{r_\uc}{\ell_{_{\rm H}}}\right)_{\rm end}^{-9}
\left(\frac{k}{aH}\right)^{-10}_\uend\right].
\end{align}
\section{Solving the CSL Equation}
The CSL equation~(\ref{eq:cslphys:app}) admits Gaussian solutions [as revealed \eg from the fact that its Lindblad counterpart~(\ref{eq:lindbladfourier}) is linear mode by mode]. Therefore, since the initial vacuum state, the Bunch-Davies state, is Gaussian, it remains so at any time and the stochastic wave function can be written as
\begin{align}
\label{eq:SingleGaussianR}
\Psi_{\bm{k}}^s\left(\eta,v_{\bm{k}}^s\right)&=
\vert N_{\bm{k}}\left(\eta\right)\vert \exp\Bigl\lbrace
-\Rea  \Omega_{\bm{k}}\left(\eta\right)
\left[v_{\bm{k}}^s-\bar{v}_{\bm{k}}^s\left(\eta\right)\right]^2
+i\sigma_{\bm k}^s(\eta)+i\chi_{\bm k}^s(\eta)
v_{\bm{k}}^s
-i\Ima  \Omega_{\bm k}(\eta )
\left(v_{\bm{k}}^s\right)^2\Bigr\rbrace\, ,
\end{align}
where, for the state to be normalised, one has
\begin{align}
\vert N_{\bm{k}}\vert =\left(\frac{2\Rea  \Omega_{\bm{k}}}{\pi}\right)^{1/4}\, .
\end{align}
In the standard picture, the quantum state
evolves into a two-mode strongly squeezed state. Here, one has
$\langle \hat{v}_{\bm{k}}^s \rangle = \bar{v}_{\bm{k}}^s$ and
$\langle \hat{p}_{\bm{k}}^s \rangle = -
i\langle \partial/\partial\hat{v}_{\bm{k}}^s \rangle = \chi_{\bm k}^s
- 2 \Ima \Omega_{\bm k}\bar{v}_{\bm{k}}^s $, giving rise to
$\langle \hat{C}^s\left(\bm{k}\right) \rangle=(\alpha_{\bm k}- 2 \Ima
\Omega_{\bm k} \beta_{\bm k})\bar{v}_{\bm{k}}^s+\beta_{\bm k}\chi_{\bm
  k}^s $.
  
For convenience, let us rewrite the CSL equation~(\ref{eq:cslphys:app}) in terms of conformal time,
\begin{align}
\label{eq:CSL:conformal}
\dd \left\vert \Psi_{\bm{k}}^s\left(\eta\right)\right\rangle &= \biggl\lbrace  
- i \hat{\mathcal{H}}_{\bm{k}}^s \dd \eta + \frac{\sqrt{ {\gamma a^4}}}{m_0} 
\left[\hat{C}^s\left(\bm{k}\right) 
- \left\langle \hat{C}^s\left(\bm{k}\right) \right\rangle \right] \dd W_\eta
-\frac{\gamma a^4}{2m_0^2} \left[\hat{C}^s\left(\bm{k}\right) 
- \left\langle \hat{C}^s\left(\bm{k}\right) \right\rangle \right]^2\dd \eta
\biggr \rbrace\left\vert \Psi_{\bm{k}}^s\left(\eta\right)\right\rangle ,
\end{align}
where
$\hat{\mathcal{H}}_{\bm{k}}^s = (\hat{p}_{\bm{k}}^s)^2/2 +
\omega^2(k,\eta) (\hat{v}_{\bm{k}}^s)^2/2 $ and where the noise
$\dd W_\eta$ is defined by $\dd W_{t}^s=a^{1/2}\dd W_\eta^s$ such that
\begin{align}
\mathbb{E}\left[\dd W_\eta^s({\bm k})\, 
\dd W_{\eta'}^{s'}({\bm k}')\right]
=\delta({\bm k}-{\bm k}')\, \delta ^{ss'}\delta(\eta-\eta')
\dd \eta^2.
\end{align}
Making use of the representation
$\hat{C}^s\left(\bm{k}\right) = \alpha_{\bm k} \hat{v}_{\bm{k}}^s -
\beta_{\bm k} i \partial/\partial\hat{v}_{\bm{k}}^s $,
the CSL equation becomes
\begin{align}
\label{eq:CSL:conformal:represented}
\frac{\dd \left\vert \Psi_{\bm{k}}^s\left(\eta\right)\right\rangle}{\dd\eta}
&=\Biggl\lbrace
-\left[\frac{i}{2}\omega^2(k,\eta) + \frac{\gamma}{2m_0^2} a^4 
\alpha_{\bm k}^2 \right] 
\left(v_{\bm{k}}^s\right)^2
+ \left(\frac{i}{2}+ \frac{\gamma}{2m_0^2} a^4 \beta_{\bm k}^2 \right) 
\frac{\partial^2}{\partial (v_{\bm{k}}^s)^2 }
+i\frac{\gamma}{m_0^2} a^4 \alpha_{\bm k} \beta_{\bm k}v_{\bm{k}}^s 
\frac{\partial}{\partial v_{\bm{k}}^s }\nonumber \\
& +\alpha_{\bm k} \left[ \frac{\sqrt{\gamma}}{m_0}a^2\frac{\dd W_\eta}
{\dd\eta}+\frac{\gamma}{m_0^2} a^4 \left(\alpha_{\bm k} 
\bar{v}_{\bm{k}}^s- 2 \Ima  
\Omega_{\bm k} \beta_{\bm k} \bar{v}_{\bm{k}}^s+\beta_{\bm k}\chi_{\bm k}^s\right)
\right] v_{\bm{k}}^s\nonumber \\
& -i\beta_{\bm k} \left[ \frac{\sqrt{\gamma}}{m_0}a^2\frac{\dd W_\eta}{\dd\eta}
+\frac{\gamma}{m_0^2} a^4 \left(\alpha_{\bm k} \bar{v}_{\bm{k}}^s
- 2 \Ima  \Omega_{\bm k} 
\beta_{\bm k} \bar{v}_{\bm{k}}^s+\beta_{\bm k}\chi_{\bm k}^s\right)\right] 
\frac{\partial}{\partial v_{\bm{k}}^s }\nonumber \\
& -\frac{\sqrt{\gamma}}{m_0}a^2\left(\alpha_{\bm k} \bar{v}_{\bm{k}}^s
- 2 \Ima  \Omega_{\bm k} 
\beta_{\bm k} \bar{v}_{\bm{k}}^s+\beta_{\bm k}\chi_{\bm k}^s\right) 
\frac{\dd W_\eta}{\dd\eta}\nonumber \\
& -\frac{\gamma}{2m_0^2}a^4 \left(\alpha_{\bm k} \bar{v}_{\bm{k}}^s
- 2 \Ima  \Omega_{\bm k} \beta_{\bm k} \bar{v}_{\bm{k}}^s
+\beta_{\bm k}\chi_{\bm k}^s\right)^2 +i\frac{\gamma}{2m_0^2}a^4 
\alpha_{\bm k} \beta_{\bm k}
\Biggr\rbrace \left\vert\Psi_{\bm{k}}^s\left(\eta\right)\right\rangle
\end{align}
Plugging \Eq{eq:SingleGaussianR} into
\Eq{eq:CSL:conformal:represented} and making use of It\^o calculus,
one can identify terms proportional to $({v_{\bm{k}}^s})^2$,
$v_{\bm{k}}^s$ and $1$. This gives rise to the set of differential
equations
\begin{align}
\frac{\dd \Rea  \Omega_{\bm{k}}}{\dd\eta} &= 
\frac{\gamma}{m_0^2} a^4\alpha_{\bm k}^2-4\frac{\gamma}{m_0^2} a^4\beta_{\bm k}^2 
\left[\left(\Rea  \Omega_{\bm{k}}\right)^2
-\left(\Ima  \Omega_{\bm{k}}\right)^2\right]
+4 \Rea  \Omega_{\bm{k}} \Ima  \Omega_{\bm{k}}
-4\frac{\gamma}{m_0^2} a^4 \alpha_{\bm k} \beta_{\bm k} \Ima  \Omega_{\bm{k}},\\
\frac{\dd \Ima  \Omega_{\bm{k}}}{\dd\eta} &= 
\frac{1}{2}\omega^2(k,\eta)-2\left[\left(\Rea  \Omega_{\bm{k}}\right)^2
-\left(\Ima  \Omega_{\bm{k}}\right)^2\right]
-8\frac{\gamma}{m_0^2} a^4 \beta_{\bm k}^2  \Rea  \Omega_{\bm{k}} 
\Ima  \Omega_{\bm{k}}+4\frac{\gamma}{m_0^2} a^4 \alpha_{\bm k} \beta_{\bm k}  
\Rea  \Omega_{\bm{k}},\\
\frac{\dd \ln\left\vert N_{\bm{k}}\left(\eta\right)\right\vert}{\dd\eta} 
& = \frac{1}{4  \Rea  \Omega_{\bm{k}}}
\frac{\dd \Rea  \Omega_{\bm{k}}}{\dd\eta},\\
\frac{\dd \bar{v}_{\bm{k}}}{\dd\eta}& = \chi_{\bm{k}}
-2\bar{v}_{\bm{k}}\Ima  \Omega_{\bm{k}}
+\frac{\sqrt{\gamma} a^2}{2m_0\Rea  \Omega_{\bm{k}}}
\left(\alpha_{\bm k} - 2 \beta_{\bm k} \Ima  \Omega_{\bm{k}}\right)
\frac{\dd W_\eta}{\dd\eta},\\
\frac{\dd\sigma_{\bm{k}}}{\dd\eta}&=-\Rea  \Omega_{\bm{k}}
+2\left(\Rea  \Omega_{\bm{k}}\right)^2\bar{v}^2_{\bm{k}}
-\frac{\chi^2_{\bm{k}}}{2}+\frac{\gamma a^4}{2m_0^2}\beta_{\bm k}
\left(\alpha_{\bm k}-2\beta_{\bm k}\Ima  \Omega_{\bm{k}}\right)
\left(1-8\Rea  \Omega_{\bm{k}} \bar{v}_{\bm{k}}^2\right)
\\ & 
-2\frac{\sqrt{\gamma}}{m_0}a^2\beta_{\bm k} \Rea  \Omega_{\bm{k}} \bar{v}_{\bm k}
\frac{\dd W_\eta}{\dd\eta}, \nonumber\\
\frac{\dd\chi_{\bm{k}}}{\dd\eta}&=2 \Ima  \Omega_{\bm{k}}\chi_{\bm{k}}
- 4 \left(\Rea  \Omega_{\bm{k}}\right)^2\bar{v}_{\bm{k}}
+8\frac{\gamma}{m_0^2} a^4\beta_{\bm k} \Rea  \Omega_{\bm{k}} \bar{v}_{\bm{k}} 
\left(\alpha_{\bm k}-2\beta_{\bm k} \Ima  \Omega_{\bm{k}}\right)
+2\frac{\sqrt{\gamma}}{m_0}a^2\beta_{\bm k} \Rea  \Omega_{\bm{k}}
\frac{\dd W_\eta}{\dd\eta}.
\end{align}
Two remarkable properties are to be noticed: $\Omega_{\bm{k}}$
decouples from the other parameters of the wavefunction, and its
dynamics is not stochastic though modified by the CSL terms. Combining
the first two above equations, one can derive an equation for
$\Omega_{\bm{k}} = \Rea \Omega_{\bm{k}} + i \Ima \Omega_{\bm{k}}$,
namely
\begin{align}
 \Omega_{\bm{k}}' = -2\left(i+2\frac{\gamma}{m_0^2} a^4 \beta_{\bm k}^2\right)
\Omega_{\bm{k}}^2+4i\frac{\gamma}{m_0^2} a^4 
\alpha_{\bm k}\beta_{\bm k} \Omega_{\bm{k}} 
+\frac{\gamma}{m_0^2} a^4\alpha_{\bm k}^2 + \frac{i}{2}\omega^2(k,\eta)\, .
\end{align}
This is a Ricatti equation that can be made linear by introducing the
function $g_{\bm k}(\eta)$ defined by the following expression
\begin{align}
\label{eq:f:g:redef}
\Omega_{\bm{k}} = \frac{1}{2\left(i+2\gamma a^4 \beta_{\bm k}^2/m_0^2\right)} 
\left(\frac{g_{\bm k}'}{g_{\bm k}}-\frac12 C_1\right)\, ,
\end{align}
and obeying 
\begin{align}
\label{eq:exactg}
g_{\bm k}''+\left(-\frac12 C_1'-\frac14C_1^2+C_2\right)g_{\bm k}=0.
\end{align}
The coefficients $C_1$ and $C_2$ are given by 
\begin{align}
\label{eq:defC12}
C_1\equiv - 2 i \frac{\gamma}{m_0^2} \left[2 a^4 \alpha_{\bm k} \beta_{\bm k}
-\frac{\left(a^4 \beta_{\bm k}^2\right)'}{1-2 i\gamma a^4 \beta_{\bm k}^2/m_0^2}
\right], 
\quad
C_2\equiv \left(1-2i\frac{\gamma}{m_0^2} a^4 \beta_{\bm k}^2\right)
\left[\omega^2(k,\eta)-2i\frac{\gamma}{m_0^2} a^4 \alpha_{\bm k}^2\right],
\end{align}
from which it follows that
$-C_1'/2-C_1^2/4+C_2=\omega^2(k,\eta)+\Delta \omega^2_\gamma
(k,\eta)$,
where $\Delta \omega^2_\gamma(k,\eta)$ is a function which vanishes
when $\gamma=0$ and can easily be determined from the expressions of
$C_1$ and $C_2$. Quite remarkably, one has
\begin{align}
\Delta \omega^2_\gamma(k,\eta)=-iS+{\cal O}\left(\gamma^2\right),
\end{align}
where $S$ is the source function introduced in \Eq{eq:source:def}, and computed in  
\Eqs{eq:sourceinf} and~(\ref{eq:sourcerad}) for inflation and radiation respectively. Solving \Eq{eq:exactg} exactly is difficult but can be done
perturbatively in $\gamma$. The perturbed solution can be written as
\begin{align}
\label{eq:expp}
g_{\bm k}(\eta)=g_{\bm k}^0(\eta)+\frac{\gamma}{m_0^2} h_{\bm k}(\eta)
+{\cal O}\left(\gamma^2\right),
\end{align}
where $g_{\bm k}^0(\eta)$ is the solution of the mode equation for
$\gamma=0$ introduced above. Plugging this expansion into \Eq{eq:exactg}, the
function $h_{\bm k}(\eta)$ obeys
\begin{align}
\label{eq:perth}
h_{\bm k}''+\omega^2(k,\eta) h_{\bm k}=i\frac{m_0^2S}{\gamma}g_{\bm k}^0,
\end{align}
which is solved as
\begin{align}
\label{eq:solh}
h_{\bm k}(\eta)=i\int _{-\infty}^\eta G(\eta,\bar{\eta}) 
\frac{m_0^2S(\bar{\eta})}{\gamma}
g_{\bm k}^0(\bar{\eta})\dd \bar{\eta},
\end{align}
where the Green function $G(\eta,\bar{\eta})$ has been introduced in \Eq{eq:Green:def}. Let us recall that the quantity $m_0^2S/\gamma $ is of order
${\cal O}(\gamma ^0)$ at leading order. Inserting the
expansion~(\ref{eq:expp}) into \Eq{eq:f:g:redef} finally leads to 
\begin{align}
\label{eq:pertomega}
\Omega_{\bm k}=\frac{1}{2i}
\frac{g_{\bm k}^0{}'}{g_{\bm k}^0}
\left\{1-\frac{\gamma}{m_0^2}\left(\frac{h_{\bm k}}{g_{\bm k}^0}
-\frac{h_{\bm k}'}{g_{\bm k}^0{}'}\right)
+i\frac{\gamma}{m_0^2}\frac{g_{\bm k}^0}{g_{\bm k}^0{}'}
\left[2a^4\alpha_{\bm k}\beta_{\bm k}
-\left(a^4\beta_{\bm k}^2\right)'\right]
+2i\frac{\gamma}{m_0^2}a^4\beta_{\bm k}^2
+{\cal O}\left(\gamma^2\right)\right\}.
\end{align}
\subsection{Inflation}
We now apply these general considerations to the case of
inflation, where the Green function is given by \Eq{eq:Green:inf} and the free source function by the expression above that equation. As already mentioned, the first term in the inflationary source
$S_\mathrm{inf}$ given in \Eq{eq:sourceinf}, \ie the one proportional to $126\epsilon_1^2$, is the dominant
one. Keeping only this term in \Eq{eq:perth}, \Eq{eq:solh} leads to
the explicit expression of $h_{\bm k}(\eta)$ which can then be used to
calculate the first correction in \Eq{eq:pertomega}. The next
step consists in calculating the two additional contributions in
\Eq{eq:pertomega}. Using the expressions of $\alpha_{\bm k}$ and
$\beta_{\bm k}$ during inflation, see \Eqs{eq:alphainf}
and~(\ref{eq:betainf}), one obtains at leading order in slow roll
$2a^4\alpha_{\bm k}\beta_{\bm k} -\left(a^4\beta_{\bm
    k}^2\right)'\simeq 108 H^2\Mp^2\epsilon_1^3/[(k\eta)^4 \eta]$
and $2a^4\beta_{\bm k}^2\simeq 36H^2\Mp^2\epsilon_1^3/(k\eta)^4$.
Inserting these results into \Eq{eq:pertomega}, one finds an exact
cancellation, meaning that it is
necessary to go to next-to-leading order in slow roll, where the result
takes the following form
\begin{align}
\Omega_{\bm k}=\Omega_{\bm k}\vert_{\gamma=0}\left[
1+4i\frac{\gamma}{m_0^2} \epsilon_1^3{\cal O}(\epsilon)
\bar{\rho}_\mathrm{inf}(-k\eta)^{-4}
+{\cal O}\left(\gamma ^2\right)\right].
\end{align}
Here, ${\cal O}(\epsilon)$ is a linear combination of the Hubble flow
parameters. Given that
$\Rea \Omega_{\bm k}\vert_{\gamma=0}=k(k\eta)^2/2$ and
$\Ima \Omega_{\bm k}\vert_{\gamma=0}=1/(2\eta)$, one finally obtains
\begin{align}
\label{eq:corr:rel:Omega:inf}
\Rea \Omega_{\bm k}=\Rea \Omega_{\bm k}\vert_{\gamma=0}
\left[1+4\frac{\gamma}{m_0^2}\epsilon_1^3{\cal O}(\epsilon)
\bar{\rho}_\mathrm{inf}(-k\eta)^{-7}\right].
\end{align}
We notice that the relative correction to $\Rea \Omega_{\bm k}$ increases with time, which is what is
needed in order
for the collapse to occur,
$\Rea \Omega_{\bm k}\gg \Rea \Omega_{\bm k}\vert_{\gamma=0}$ . If one requires the collapse to happen
during inflation, a lower bound on the parameter
$\gamma ,$ defined to be its value such that the relative correction evaluated
at $\eta_\uend$ is larger than one, can be placed. Of course, this limit depends on
the unknown factor ${\cal O}(\epsilon)$. However, as discussed below, the collapse is more efficient during the radiation-dominated era, and the precise value of that quantity plays no role.
\subsection{Radiation dominated epoch}
During radiation, the Green function is given by \Eq{eq:Green:rad} and the free mode function by \Eq{eq:free:mode:rad}. Using the expressions of $\alpha_{\bm k}$ and
$\beta_{\bm k}$ during the radiation-dominated era, one also has, for the last two terms in \Eq{eq:pertomega},
$2a^4\alpha_{\bm k}\beta_{\bm k} -\left(a^4\beta_{\bm
    k}^2\right)'\simeq 864\eta_\uend^4
H_\uend^2\Mp^2\left[3(\eta-\eta_\mathrm{r})^2
  -k^2\eta_\uend^4H_\uend^2r_\uc^2\right]/[k^4(\eta-\eta_\mathrm{r})^{11}]$ and $2a^4\beta_{\bm k}^2\simeq 864 \eta_\uend^4
H_\uend^2\Mp^2/[k^4(\eta-\eta_\mathrm{r})^8]$.
\subsubsection{Case where the mode crosses out $r_\uc$ during inflation}
As explained above, the first term in \Eq{eq:sourcerad} for $S_\mathrm{rad}$ is the dominant one in that case, and at leading order in $r_\uc/\lambda_\uend$, one obtains
\begin{align}
\Omega_{\bm k}&\simeq 
\Omega_{\bm k}\vert_{\gamma=0}
\biggl[1
+i\frac{\gamma}{m_0^2}
 1152 \frac{\bar{\rho}_\uend}{k^4(-\eta_\uend)^3(\eta-\eta_\mathrm{r})}
+{\cal O}\left(\gamma^2\right)\biggr].
\end{align}
Given that
$\Rea \Omega_{\bm
  k}\vert_{\gamma=0}=(k\eta_\uend)^4/[2k(\eta-\eta_\mathrm{r})^2]$
and
$\Ima \Omega_{\bm k}\vert_{\bm k}=-[2(\eta-\eta_\mathrm{r})]^{-1}$, we
notice that the correction has the same time dependence as
$\Rea \Omega_{\bm k}\vert_{\gamma=0}$, so its relative value is frozen to
\begin{align}
\Rea \Omega_{\bm k}\simeq \Rea \Omega_{\bm k}\vert_{\gamma=0}
\left[1+1152\frac{\gamma}{m_0^2}\bar{\rho}_\uend(-k\eta_\uend)^{-7}
+{\cal O}\left(\gamma^2\right)\right].
\end{align}
This correction is larger than in \Eq{eq:corr:rel:Omega:inf}, which justifies the statement that the collapse is more efficient in the radiation-domiated epoch. The condition for the collapse, \ie having a relative correction of
order one, is then
\begin{align}
\frac{\gamma}{m_0^2}>(1152\bar{\rho}_\uend)^{-1}(-k\eta_\uend)^7.
\end{align}
\subsubsection{Case where the mode crosses out $r_\uc$ during the radiation-dominated era}
As already
discussed, three terms must be kept in the
expansion~(\ref{eq:sourcerad}) of $S_\mathrm{rad}$, namely the terms
proportional to the coefficients $3024$, $216$ and $1836$. This gives rise to
\begin{align}
\Omega_{\bm k}&\simeq 
\Omega_{\bm k}\vert_{\gamma=0}
\biggl[1
+i\frac{\gamma}{m_0^2} \frac{21792}{11}\frac{H_\uend^2\Mp^2}{k(-k\eta_\uend)^{10}
(H_\uend r_\uc)^7(\eta-\eta_\mathrm{r})}
-i\frac{\gamma}{m_0^2}864\eta_\uend^4
H_\uend^2\Mp^2\frac{3(\eta-\eta_\mathrm{r})^2
  -k^2\eta_\uend^4H_\uend^2r_\uc^2}{k^4(\eta-\eta_\mathrm{r})^{12}}
\nonumber \\ &
+i\frac{\gamma}{m_0^2}\frac{864 \eta_\uend^4
H_\uend^2\Mp^2}{k^4(\eta-\eta_\mathrm{r})^8}
+{\cal O}\left(\gamma^2\right)\biggr].
\end{align}
We see that the two last terms are subdominant. In this approximation, 
the relative correction is again time-independent and given by 
\begin{align}
\Rea \Omega_{\bm k}\simeq \Rea \Omega_{\bm k}\vert_{\gamma=0}
\left[1+\frac{7264}{11}\frac{\gamma}{m_0^2}
\bar{\rho}_\uend (k\eta_\uend)^{-14}(H_\uend r_\uc)^{-7}
+{\cal O}\left(\gamma^2\right)\right].
\end{align}
The lower bound on the parameter $\gamma$ can therefore be expressed
as
\begin{align}
\frac{\gamma}{m_0^2}>\left(\frac{7264}{11}\bar{\rho}_\uend\right)^{-1}
(-k\eta_\uend)^{14}
(H_\uend r_\uc)^7.
\end{align}

\section{Density Contrasts and the Parameter $p$}

In the theory of cosmological perturbations, in the scalar sector, the
most general perturbed metric tensor reads
\begin{align}
\label{eq:ds2}
\dd s^2 = a^2(\eta )\left\{- \left(1+2\phi\right)\dd \eta ^2 + 2\partial _iB
\dd x^i \dd \eta + \left[\left(1-2\psi\right)\delta_{ij}+2\partial_i\partial _jE
\right]\dd x^i{\rm d}x^j\right\} ,
\end{align}
where $\phi$, $B$, $\psi$ and $E$ are four scalar functions of space
and time. As is well-known, the theory features a ``gauge symmetry'',
meaning that the quantities appearing in \Eq{eq:ds2} are in general
not invariant under (small) space-time diffeomorphisms and, therefore,
cannot be considered as observables. The cure is then either to
specify a particular system of coordinates or to work in terms of
``gauge-invariant'' quantities, that is to say quantities that are
invariant under a small change of coordinates. The most general change
of coordinates that can be constructed with scalar functions [given
  here by the scalar functions $\xi ^0(\eta,{\bm x})$ and
  $\xi(\eta,{\bm x})$] is
\begin{align}
  \label{eq:coord}
\eta \rightarrow \tilde{\eta }=\eta +\xi ^0\left(\eta ,{\bm x}\right),\quad 
x^i \rightarrow \tilde{x}^i=x^i
+\delta ^{ij}\partial _j\xi \left(\eta ,{\bm x}\right)\, .
\end{align}
Then, we find that the four scalar functions used to construct the
scalar perturbed metric given by \Eq{eq:ds2} transform, under the
above change of coordinates~(\ref{eq:coord}), according to
\begin{align}
\label{eq:liescalar}
\tilde{\phi}=\phi +\xi ^{0'}+\frac{a'}{a}\xi ^0, \quad 
\tilde{B}=B-\xi ^0+\xi ', \quad \tilde{\psi }=\psi -\frac{a'}{a}\xi ^0, 
\quad \tilde{E}=E+\xi \, .
\end{align}
As a consequence, if we now consider the two following combinations
\begin{align}
\Phi \equiv \phi +\frac{1}{a}\left[a\left(B-E'\right)\right]',\quad 
\Psi \equiv \psi  -\frac{a'}{a}\left(B-E'\right)\, ,
\end{align}
then it is easy to establish that these two quantites are
gauge-invariant: $\tilde{\Phi }=\Phi$ and $\tilde{\Psi }=\Psi $. They
are called the Bardeen potentials~\cite{Bardeen:1980kt}.

Of course, for consistency, the stress-energy tensor describing matter
must also be perturbatively expanded and, as a consequence, one needs to
construct gauge-invariant combinations for the scalar quantities
appearing in $\delta T_{\mu \nu}$, in particular for the density
contrast. From the rule of transformation of two-rank tensors, one
obtains
\begin{align}
\tilde{\delta }=\delta +\frac{\rho '}{\rho }\, \xi ^0\,, \quad 
\tilde{v}=v-\xi '\,, \quad
\tilde{\delta p}=\delta p+p'\xi ^0\, ,
\end{align}
where $\delta\equiv \delta \rho/\rho$ is the density contrast, $v$ the
peculiar velocity and $\delta p$ the perturbed pressure. As is
well-known, it is possible to build various density contrasts that are
gauge invariant. Two proto-typical examples are given by
\begin{align}
\delta _\mathrm{g}\equiv \delta +\frac{\rho '}{\rho }(B-E')\, ,\quad 
\delta _\mathrm{m} \equiv \delta +\frac{\rho '}{\rho }(v+B)\, .
\end{align}
More generally, a gauge-invariant density contrast can always be
intoduced by considering the following definition
\begin{align}
  \delta_\mathrm{inv}=\delta +\frac{\rho'}{\rho}{\cal D}(\phi,\psi,B,E),
\end{align}
where ${\cal D}$ is an arbitrary function of $\phi$, $\psi$, $B$ and
$E$ and their derivatives, provided it satisfies $\tilde{\cal
  D}\rightarrow {\cal D}-\xi^0$. One easily checks that this is the
case for ${\cal D}_\mathrm{g}=B-E'$ or ${\cal
  D}_\mathrm{m}=v+B$. Using the fact that the behaviour of $\delta
_{\rm g}$ is given by the time-time Einstein equation, namely $\delta
_{\rm g}=-2\Mp^2[3{\cal H}({\cal H}\Phi +\Phi ')+k^2\Phi ]/(\rho
a^2)$, one can also write
\begin{align}
  \label{eq:deltainv}
  \delta_\mathrm{inv}=-\frac{2\Mp^2}{\rho a^2}
        \left[3{\cal H}({\cal H}\Phi +\Phi ')+k^2\Phi \right]
        +\frac{\rho'}{\rho}\left[{\cal D}(\phi,\psi,B,E)-B+E'\right].
\end{align}
We are then interested in the limit $k\rightarrow 0$, namely the large-scale 
limit for which different situations can occur. The most generic
one is that ${\cal D}$ is a function of $\phi$, $\psi$, $B$ and $E$
where positive powers of $k$ appear. In the large-scale limit, the
scale-dependent terms will be negligible and the scale dependence of
$\delta _\mathrm{inv}$ will be that of $\delta _\mathrm{g}$, namely
$p=0$ for the parameter $p$ introduced in the main text. Of course, one
exception is when ${\cal D}$ is such that it cancels out all the
scale-independent terms in \Eq{eq:deltainv}, leaving $k^2\Phi$ as the
leading term. This case is nothing but
$\delta_\mathrm{inv}=\delta_\mathrm{m}$ and corresponds to
$p=2$. Clearly, this case exists but is ``fine-tuned''. Note that the only way to modify these conclusions is to incorporate negative powers of $k$ in ${\cal D}$. However, this would correspond to having a non-local function in real space, which is not very realistic. In brief,
we have shown that either $p=0$ or $p=2$, with this last case being in some sense of ``zero measure''.

\end{widetext}

\bibliography{CSL}

\end{document}